\documentclass[%
      aps,
      prb, 
      reprint,
      twocolumn,
      floatfix
      superscriptaddress,
      showpacs
      ]{revtex4-2}
\usepackage[english]{babel}
\usepackage{microtype}
\usepackage{indentfirst}
\usepackage{graphicx}
\usepackage{xcolor}
\usepackage{booktabs}
\usepackage{tabularx}
\usepackage{multirow}
\usepackage{braket}
\usepackage{amsmath}
\usepackage{nicematrix} 
\usepackage{dsfont}
\usepackage{array}
\usepackage{dcolumn}
\usepackage{bm}
\usepackage{setspace}
\usepackage[normalem]{ulem}
\usepackage{color}
\usepackage[varg]{txfonts}
\usepackage[small]{titlesec}

\usepackage[ colorlinks,
    linkcolor={blue},
    citecolor={blue},
    urlcolor={blue}]{hyperref}

\usepackage{tikz}

\definecolor{maxgreen}{rgb}{0,0.75,0}
\definecolor{maxviolet}{rgb}{0.52,0,0.56}
\definecolor{maxorange}{rgb}{1,0.61,0}
\definecolor{maxred}{rgb}{0.62,0,0}
\definecolor{maxblue}{rgb}{0,0.45,0.74}
\definecolor{maxgrey}{rgb}{0.62,0.62,0.62}

\definecolor{darkbrown}{HTML}{a6611a}
\definecolor{lightbrown}{HTML}{dfc27d}
\definecolor{lightmint}{HTML}{80cdc1}
\definecolor{darkmint}{HTML}{018571}

\newcommand{\reviewin}[1]{{\color{black}#1}}
\newcommand{\reviewout}[1]{}

\newcommand{\br}{\mathbf{r}}
\newcommand{\bre}{\br_e}
\newcommand{\brh}{\br_h}
\newcommand{\Xian}{{X}_{ia}^{(n)}}
\newcommand{\Yian}{{Y}_{ia}^{(n)}}
\newcommand{\Xjan}{{X}_{ja}^{(n)}}
\newcommand{\Yjan}{{Y}_{ja}^{(n)}}
\newcommand{\Xibn}{{X}_{ib}^{(n)}}
\newcommand{\Yibn}{{Y}_{ib}^{(n)}}
\newcommand{\Xjbn}{{X}_{jb}^{(n)}}
\newcommand{\Yjbn}{{Y}_{jb}^{(n)}}
\newcommand{\Psiexc}{\Psi_\mathrm{exc}^{(n)}}
\newcommand{\Omexc}{\Omega^{(n)}}
\newcommand{\bX}{\mathbf{X}}
\newcommand{\bY}{\mathbf{Y}}

\newcommand{\evgwnod}{ {\text{ev}GW_0} }
\newcommand{\exc}{_\text{exc}^{(n)}}

\newcommand{\li}{ \underset{L \rightarrow \infty}{\lim}}

\begin{document}

\title{Optical excitations in nanographenes from the Bethe-Salpeter equation and time-dependent density functional theory: absorption spectra and spatial descriptors}

\author{Maximilian Graml}\email{maximilian.graml@physik.uni-regensburg.de}
\author{Jan Wilhelm}
\affiliation{Institute of Theoretical Physics and Regensburg Center for Ultrafast Nanoscopy (RUN), University of Regensburg, 93053 Regensburg, Germany}
\date{\today}
\begin{abstract}
The \textit{GW} plus Bethe-Salpeter equation (\textit{GW}-BSE) formalism is a well-established approach for calculating excitation energies and optical spectra of molecules, nanostructures, and crystalline materials.
We implement \textit{GW}-BSE in the CP2K code and validate the implementation for a standard organic molecular test set, obtaining excellent agreement with reference data, with a mean absolute error in excitation energies below 3~meV. 
We then study optical spectra of nanographenes of increasing length,  showing excellent agreement with experiment. 
We further compute the size of the excitation of the lowest optically active excitation which converges to about 7.6~{\AA} with increasing length. 
Comparison with time-dependent density functional theory using functionals of varying exact-exchange fraction shows that none reproduce both the size of the excitation and optical spectra of \textit{GW}-BSE, underscoring the need for many-body methods for accurate description of electronic excitations in nanostructures.

\end{abstract}

\maketitle

\section{Introduction}
When light interacts with a semiconductor, it can promote an electron to the conduction band, leaving behind a hole in the valence band. 
The electron and hole attract each other through the Coulomb interaction and can form a bound pair known as an exciton~\cite{Ullrich2011, Onida2002, Martin2016, Kunstmann2018, Merkl2019, Siday2022, Naik2022, Schmitt2022, Bange2024, Liebich2025}. 
A correct description of such a bound state requires including the screened Coulomb interaction between the electron and the hole. 
Time-dependent density functional theory (TDDFT) with local or semi-local exchange-correlation functionals is widely used to compute excitation energies, but in these approximations the long-range electron-hole attraction via Coulomb interaction is absent, and bound excitons are not described~\cite{Onida2002, Ullrich2011, Martin2016}. 
Hybrid functionals include a fraction of nonlocal Hartree-Fock exact exchange, which partially restores the bare Coulomb interaction between the electron and hole. 
For example, a hybrid with 25\% exact exchange corresponds, in a simple picture, to an effective static dielectric screening of about four. 
While hybrids can improve agreement with experiment, their accuracy depends sensitively on the chosen exact-exchange fraction and they are significantly more computationally expensive than local or semi-local functionals. 

An alternative approach is the $GW$ plus Bethe-Salpeter equation ($GW$-BSE)~\cite{Strinati1988, Rohlfing2000, Onida2002, Bechstedt2015, Martin2016}, which explicitly includes the screened Coulomb attraction between an excited electron and the hole it leaves behind. 
The $GW$-BSE framework has become a standard in condensed matter physics for computing optical spectra of semiconductors~\cite{Strinati1988, Benedict1998, Albrecht1998, Rohlfing2000, Onida2002, Marini2009, Deslippe2012, Qiu2013, Bradley2015, Sander2015, Bechstedt2015, Martin2016, Vorwerk2019, Sangalli2019, Merkel2023, Merkel2023a, Zhou2025}, and is also used in chemistry for accurate calculations of molecular excitation energies~\cite{Jacquemin2015, Jacquemin2015a, Bruneval2015, Bruneval2016,Blase2016,Jacquemin2016, Rangel2017, Jacquemin2017, Krause2017, Gui2018, Blase2018, Holzer2018a, Blase2020, Liu2020, Forster2022, Rauwolf2024, Knysh2024, Holzer2025}, where a relation to the established wavefunction-based approaches has been shown~\cite{Scuseria2008, Berkelbach2018, Lange2018, Quintero-Monsebaiz2022, Tolle2023}.
\reviewin{The intermediate regime between small molecules and crystals include nanometer-sized materials, like nanographenes which are finite cutouts of a graphene with hydrogen termination that can be synthesized on noble metal surfaces~\cite{Cai2010,  Jiang2023}.
Applications of them include nanoscale transistors~\cite{Borin2022, Zhang2026} and nanoscale optoelectronic devices~\cite{Denk2014}, where $GW$-BSE has been used to characterize the optical absorption spectrum.
Another class of nanoscale materials are moiré superlattices build from twisted van der Waals crystals, where $GW$-BSE provides direct access to both the excitation energies and the spatial structure of the electron-hole pair~\cite{Naik2022}. 
Here,}
%
%
electron and hole can separate over different layers~\cite{Kunstmann2018} ("interlayer exciton"), or can be located within a single layer ("intralayer exciton")~\cite{Naik2022}, where hybrid states~\cite{Siday2022} as well as the switching between distinct intra-/interlayer character have been observed experimentally~\cite{Merkl2019,Schmitt2022,Bange2024,Liebich2025}.
The spatial structure of the electron-hole pairs determines their excitation energy and the spatial structure is thus linked to the peak position in optical spectra. 

In principle, the spatial properties of an excited state can be obtained directly from its wavefunction $\Psiexc(\bre,\brh)$, which depends on the electron coordinate $\bre$ and the hole coordinate $\brh$. 
A common way to visualize $\Psiexc(\bre,\brh)$ is to fix one coordinate and plot the probability density of the other, for example $|\Psiexc(\bre,\brh^*)|^2$ for a chosen hole position $\brh^*$~\cite{Naik2022}. 
The result depends on the choice of $\brh^*$, and important features may be overlooked if the chosen reference position is not representative. 
As an alternative, following the ideas of Frenkel~\cite{Frenkel1931, Frenkel1931a} and Wannier~\cite{Wannier1937} in distinguishing between different types of excitons, one can characterize the excited state by expectation values of the electron-hole separation, the electron position, and the hole position. 
These spatial descriptors have been introduced in quantum chemistry to study electronic excitations in molecules and polymer chains~\cite{Plasser2012, Plasser2014a, Plasser2014b, Bappler2014, Plasser2015, Mewes2015, Mewes2016, Mewes2017, Hirose2017, Mewes2018, Mewes2019, Kimber2020}. 
The same expectation values can be evaluated as functions of time, providing a direct and compact description of the spatio-temporal evolution of an excited state. 
Such time-dependent descriptors can be computed using time-dependent $GW$-BSE~\cite{Attaccalite2011, Jiang2021, Sangalli2021, Chan2021, Perfetto2022, Marek2025} or real-time TDDFT~\cite{Ullrich2011, Li2020, Mattiat2022, Kick2024, Schreder2024, Choi2024, Hanasaki2025}. 
Beyond the time evolution of electronic states, recent advances in force calculations within $GW$-BSE~\cite{Villalobos-Castro2023, Tolle2025, Tolle2025a, Kitsaras2026} enable the description of coupled electronic and nuclear motion\reviewin{, where a simplified frequency treatment in the self-energy appears promising to reduce computational cost and to ensure numerical stability~\cite{Berger2021,Tyagi2024,Tyagi2025,Tyagi2025a}. }
Recent experimental advances enable tracking of such excited-state dynamics with simultaneous femtosecond temporal and sub-nanometer spatial resolution~\cite{Schmitt2022, Bange2024, Roelcke2024, Siday2024, Zizlsperger2024, Anglhuber2025, Maier2025}, allowing direct comparison between measured quantities such as the time-dependent exciton radius and their theoretical predictions. 
The values of spatial descriptors for excited states, such as the size of the excitation, depend on the method used to compute the excitation, for example TDDFT or $GW$-BSE. 
The aim of this work is to assess how the choice of the excited-state method influences these descriptors in finite nanostructures. 
To this end, we implement the $GW$-BSE formalism in the CP2K software~\cite{CP2K,Kuhne2020,Iannuzzi2026}, enabling the study of both optical excitations and their spatial extent in nanostructures. 
We first present the theory of the $GW$-BSE framework in Sec.~\ref{sec-theory}. 
$GW$-BSE calculations on  nanographenes are presented in Sec.~\ref{sec-spectra_of_nanographenes_theory}, where we analyze the convergence of the optical absorption spectrum with increasing size and compare the computed absorption spectrum to experimental data. 
In Sec.~\ref{sec-theory_of_descriptors}, we introduce the theory of spatial descriptors and we apply these to nanographenes in Sec.~\ref{sec-spatial_descriptors}, including a comparison to TDDFT.

\section{Optical absorption spectrum from \textit{GW}-BSE}\label{sec-theory}

\begin{figure}[t!]
    \centering
    \includegraphics[width=\linewidth]{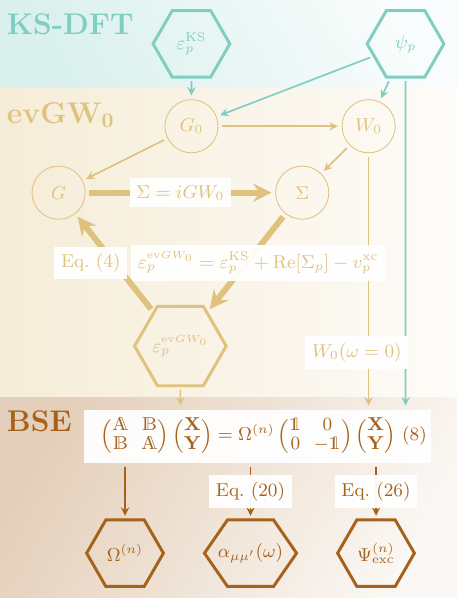}
    \caption{
    Workflow of a \textit{GW}-BSE calculation within this work, where we employ BSE@ev$GW_0$@PBE.
    A KS–DFT calculation provides the KS energies $\varepsilon_p^{\mathrm{KS}}$ and orbitals $\psi_p^{\mathrm{KS}}$ as input.  
    In the ev$GW_0$ scheme, the quasiparticle energies are iterated (bold arrows) in a self-consistent loop, starting from $G=G_0$ and using $W_0$ from the DFT starting point.  
    This yields the ev$GW_0$ quasiparticle energies $\varepsilon_p^{\mathrm{ev}GW_0}$ [Eq.~\eqref{eq-evGW0_energies}].  
    BSE matrices~$\mathds{A},\mathds{B}$ [Eq.~\eqref{eq-BSE_ingredients}] are  constructed from $\varepsilon_p^{\mathrm{ev}GW_0}$, $\psi_p^{\mathrm{KS}}$, and the statically screened Coulomb interaction $W_0(\omega=0)$.  
    Diagonalizing Eq.~\eqref{eq-BSE_ABBA} gives the excitation energies $\Omega^{(n)}$, the absorption spectrum $\alpha_{\mu\mu'}(\omega)$, and the excitation wavefunction $\Psi^{(n)}_{\mathrm{exc}}(\mathbf{r}_e,\mathbf{r}_h)$.
    }
    \label{fig-flowchart}
\end{figure}

The \textit{GW}-BSE formalism provides an accurate description of neutral excitations by explicitly accounting for the electron-hole interaction on top of a quasiparticle picture~\cite{Strinati1988, Rohlfing2000, Onida2002, Bechstedt2015, Martin2016}. 
Starting from Kohn-Sham (KS) density functional theory (DFT), we first obtain the ground-state electronic structure, which provides a convenient single-particle basis~$\{\psi_p(\br)\}$ for many-body perturbation theory. 
The KS energies $\varepsilon_p^{\mathrm{KS}}$ and orbitals $\psi_p(\br)$ are determined by solving
\begin{align}
    [h_0(\br) + v^\mathrm{xc}(\br) ]\psi_p(\br) & = \varepsilon_p^{\mathrm{KS}} \psi_p(\br) \, , \label{eq-KS_equation}
\end{align}
where $h_0(\br)$ contains the kinetic energy, the external potential, and the Hartree potential, and $v^\mathrm{xc}(\br)$ is the exchange-correlation potential. 
In our implementation, the KS orbitals are expanded in Gaussian basis functions $\{\phi_\nu(\br)\}$ as
\begin{align}
    \psi_p(\br) = \sum_\nu C_{\nu p}\phi_\nu(\br) \, , \label{eq-basis_set_expansion_KS_orb}
\end{align}
with molecular orbital indices $p,q,r,s$, atomic orbital indices $\nu$, and we can explicitly separate KS states~$\psi_p$ into $N_\text{occ}$ occupied states~$\psi_i,\psi_j$ and $N_\text{empty}$ unoccupied states $\psi_a,\psi_b$. 
These KS states serve as the starting point for the $GW$ calculation (cf. Fig.~\ref{fig-flowchart}), in which the poles of the single-particle Green's function $G$ can be interpreted as quasiparticle energies, effectively improving the KS energies $\varepsilon_p^\text{KS}$.
In practice, we apply a simplification of Hedin's equations~\cite{Hedin1965} \reviewin{by replacing the vertex with a single spacetime point (cf. App.~\ref{app-gw_schemes})}, which leads to a self-energy of the form 
\begin{align}
    \Sigma^{GW}(\br_1, \br_2, t) = 
    i G(\br_1, \br_2, t) W(\br_1, \br_2, t) \, . \label{eq-GW}
\end{align}
Hence, this simplification is called the $GW$ approximation, where $W$ denotes the dynamically screened Coulomb interaction. 
\reviewin{In practice, the resulting equations are usually solved non-selfconsistently, various flavors are summarized in App.~\ref{app-gw_schemes}.
Here, we use eigenvalue self-consistency in $G$, i.e.~ev$GW_0$.}
\reviewout{There,}\reviewin{In ev$GW_0$}, the KS energies $\varepsilon_p^\text{KS}$ and KS orbitals $\psi_p(\br)$ enter the non-interacting Green's function~\cite{Wilhelm2016}\footnote{Here, we adapt the typically used nomenclature, although our initial guess from KS-DFT should rather be called 'mean-field' $G_0$~\cite{Golze2019}.}
\begin{align}
    G_0(\br_1,\br_2,i\omega) = \sum_p \frac{\psi_p(\br_1)\psi_p(\br_2)}{i\omega + \varepsilon_F - \varepsilon_p^\text{KS}} \, , \label{eq-G0}
\end{align}
with the Fermi energy $\varepsilon_F$.
Additionally, the screened Coulomb interaction $W_0$ is computed beforehand as~\cite{Wilhelm2016}
\begin{align}
    W_0(\br_1,\br_2,i\omega) = \int \! \mathrm{d}^3r' \epsilon^{-1}(\br_1,\br', i\omega) v(\br',\br_2) \, , \label{eq-W0}
\end{align}
where $v(\br_1,\br_2) = 1/|\br_1 - \br_2|$ is the bare Coulomb interaction and the dynamical dielectric function $\epsilon(\br_1,\br', i\omega)$ is computed at the RPA level from the KS energies $\varepsilon_p^\text{KS}$ \reviewin{(cf. App.~\ref{app-gw_schemes})}.
Formally, Eq.~\eqref{eq-W0} can be rewritten using a Green's function~\cite{Golze2019}, as depicted for $G_0$ in Fig.~\ref{fig-flowchart}.
Combining Eqs.~\eqref{eq-G0} and \eqref{eq-W0} with Eq.~\eqref{eq-GW}, we can compute the quasiparticle energies from
\begin{align}
    \varepsilon_p^{\evgwnod} = \varepsilon_p^\text{KS} + \mathrm{Re}[\Sigma^{\evgwnod}_p(\varepsilon_p^{\evgwnod})] - v_p^\text{xc} \, . \label{eq-evGW0_energies}
\end{align}
With the solution of this non-linear equation, we recompute the Green's function~\eqref{eq-G0} by replacing $\varepsilon_p^\text{KS}$ with $\varepsilon_p^{\evgwnod}$~\cite{Golze2019}:
\begin{align}
    G_\text{ev$GW_0$}(\br_1,\br_2,i\omega) = \sum_p \frac{\psi_p(\br_1)\psi_p(\br_2)}{i\omega + \varepsilon_F - \varepsilon_p^{\evgwnod}} \, , \label{eq-G}
\end{align}
We repeat this procedure until convergence in $\varepsilon_p^{\evgwnod}$ is reached, as depicted by the triangle in the ev$GW_0$-box in Fig.~\ref{fig-flowchart}.

While the poles of the single-particle Green's function provide the (formally) correct quasiparticle energies, optical (two-particle) excitations need an extended framework, which includes two-particle Green's function.
This can be done by reintroducing the previously neglected vertex corrections~\cite{Onida2002} in Hedin's equations~\cite{Hedin1965}, which eventually leads to the BSE for the four-point polarizability~\cite{Strinati1988,Rohlfing2000,Onida2002}. 
In App.~\ref{app-derivation_dyn_dip_pol_and_abba_eq}, we derive how the solution of this rather complicated form of the BSE can be transformed to the following generalized eigenvalue problem in the product space of occupied and unoccupied orbitals, which is then typically solved to describe optical excitations~\cite{Onida2002, Rohlfing2000,Jacquemin2016, Bechstedt2015, Liu2020, Bruneval2015, Bruneval2016, Blase2020}:
\begin{align}
    \left( \begin{array}{cc}\mathds{A} &  \mathds{B}\\\mathds{B} &  \mathds{A}\end{array} \right)\left( \begin{array}{cc}\bX^{(n)}\\\bY^{(n)}\end{array} \right) = \Omexc\left(\begin{array}{cc}\mathds{1}&0\\0&-\mathds{1}\end{array}\right)\left(\begin{array}{cc}\bX^{(n)}\\\bY^{(n)}\end{array}\right)  \, , \label{eq-BSE_ABBA}
\end{align}
where $\Omexc$ denotes the energy of excitation $n$ and $\bX^{(n)}$ and $\bY^{(n)}$ are bi-orthogonal eigenvectors following~\cite{Thouless1961, McLACHLAN1964, Ullah1971, Oddershede1984,Stratmann1998,Ullrich2011, Sander2015, Plasser2025}
\begin{align}
    \sum_{i,a} X_{ia}^{(m)} X_{ia}^{(n)} - Y_{ia}^{(m)} Y_{ia}^{(n)} = \pm \delta_{mn} \, . \label{eq-normalization_XY}
\end{align}
The sign corresponds to positive and negative excitation energies $\Omexc$, respectively, as the solutions of Eq.~\eqref{eq-BSE_ABBA} come in pairs of ($\Omexc,\bX^{(n)},\bY^{(n)}$) and ($-\Omexc,\bY^{(n)},\bX^{(n)}$)~\cite{Thouless1961, McLACHLAN1964, Ullah1971, Ullrich2011, Plasser2025} with excitation index $n=1,\dots,N_\text{occ}N_\text{empty}$.
We choose $\Xian$ and $\Yian$ to be real-valued for the finite structures under study.

For a closed-shell ground state, the block matrices in Eq.~\eqref{eq-BSE_ABBA} are defined by (cf. App.~\ref{app-derivation_dyn_dip_pol_and_abba_eq})~\cite{ Ljungberg2015, Jacquemin2016}
\begin{align}
    A_{ia,jb} &= (\varepsilon_a^{\evgwnod}-\varepsilon_i^{\evgwnod})\delta_{ij}\delta_{ab} + \alpha^\mathrm{(S/T)}
    v_{ia,jb} - W_{ij,ab} \,, \nonumber
    \\
    B_{ia,jb} &= \alpha^\mathrm{(S/T)} v_{ia,bj} - W_{ib,aj} \,,
    \label{eq-BSE_ingredients}
\end{align}
where the preceding KS-DFT and $GW$ calculations (cf. Fig.~\ref{fig-flowchart}) provide quasiparticle energies $\varepsilon_p^{\evgwnod}$, and 
\begin{align}
    v_{pq,rs} &= \!\int \! \text{d}^3r \, \text{d}^3r' \psi_p(\br) \psi_q(\br) v(\br,\br') \psi_r(\br') \psi_s(\br') \label{eq-components_unscreened_coulomb}
\\
    W_{pq,rs} &=\hspace{-0.3em} \int \! \text{d}^3r \, \text{d}^3r' \psi_p(\br) \psi_q(\br) W_0(\br,\br'\hspace{-0.25em}, \omega{=}0) \psi_r(\br') \psi_s(\br') \label{eq-components_screened_coulomb}
\end{align}
are matrix elements of the bare and the statically RPA-screened Coulomb interaction, respectively.
The prefactor $\alpha^\text{(S/T)}$ depends on the spin configuration and is $\alpha^\text{(S)} = 2$ for a singlet excited state and $\alpha^\text{(T)} = 0$ for a triplet excited state (cf. App.~\ref{app-derivation_dyn_dip_pol_and_abba_eq}).
In practice, we compute the matrix elements of Eqs.~\eqref{eq-components_unscreened_coulomb}/\eqref{eq-components_screened_coulomb} using the resolution-of-the-identity (RI) technique~\cite{Liu2020, Krause2017}, where auxiliary RI basis functions $\varphi_{P}$ (indexed by $P,Q,R$) are introduced:
\begin{align}
    v_{pq,rs} &\approx \sum_P B_{pq}^P B_{rs}^P \, ,\\
    W_{pq,rs} &\approx \sum_{P} B_{pq}^P (\epsilon^{-1})_{PQ} B_{rs}^P \, .
\end{align}
Here, the coefficients $B_{pq}^P$ are computed from~\cite{Wilhelm2016}
\begin{align}
    B_{pq}^P = \sum_Q (pq|Q) \ L_{PQ}^{-1} \, ,
\end{align}
where $(pq|Q)$ denotes three-center Coulomb integrals
\begin{align}
    (pq|Q) = \int \! \text{d}^3r \text{d}^3r' \psi_p(\br') \psi_q(\br') v(\br,\br') \varphi_Q(\br) \, ,\label{eq-3c}
\end{align}
and the $L_{PQ}$ are obtained from a Cholesky decomposition of the Coulomb matrix
\begin{align}
    (P|Q) = \int \! \text{d}^3r \ \varphi_P(\br) v(\br,\br') \varphi_Q(\br') = \sum_R L_{PR} L_{RQ}^T \, .\label{eq-2c}
\end{align}
Note that for the Gaussian basis~\eqref{eq-basis_set_expansion_KS_orb} employed in this work, analytical expressions for two- and three-center Coulomb integrals~\eqref{eq-3c}, \eqref{eq-2c} guarantee efficient computations~\cite{Obara1986,Libint,Golze2017}.  
When solving Eq.~\eqref{eq-BSE_ABBA} in practice, we assume that $ \mathds{A}{-} \mathds{B}$ is positive definite, which allows us to recast Eq.~\eqref{eq-BSE_ABBA} into a hermitian eigenvalue problem with half the size. 
We skip the details of this procedure and refer instead to the existing literature~\cite{Ullah1971, McCurdy1971, Jorgensen1981, Oddershede1984, Stratmann1998, Furche2001, Ullrich2011, Sander2015, Bruneval2015, Herbert2023}.
By neglecting the coupling blocks $ \mathds{B}$, Eq.~\eqref{eq-BSE_ABBA} can be simplified even further, which is called the Tamm-Dancoff approximation (TDA)~\cite{Benedict1998}:
\begin{align}
    \mathds{A} \bX_\text{TDA}^{(n)} = \Omexc_\text{TDA} \bX_\text{TDA}^{(n)} \label{eq-TDA_of_BSE}
\end{align}
Computationally, both approaches are  demanding in memory  ($\mathcal{O}(N_\mathrm{occ}^2N_\mathrm{empty}^2)$) and in the number of floating point operations for solving the eigenvalue problem ($\mathcal{O}(N_\mathrm{occ}^3N_\mathrm{empty}^3)$).
Correspondingly, in order to enable the solution of Eq.~\eqref{eq-BSE_ABBA} for large-scale systems, we adopt the standard procedure and reduce the associated prefactors by truncating the considered number of orbitals in Eq.~\eqref{eq-BSE_ABBA} by an energy cutoff~\cite{Liu2020}.
In App.~\ref{app-cutoff_checks}, we discuss the details of this technique and to which extent it affects the accuracy of the optical absorption spectrum and spatial properties of the excitations.

\begin{figure}[t!]
    \centering
    \includegraphics[width=3.375in]{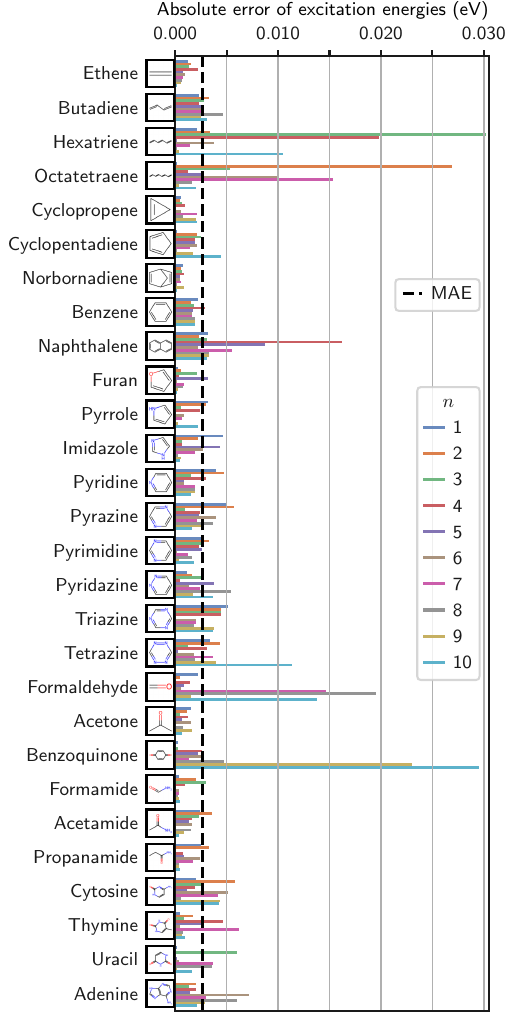}
    \caption{Absolute error~$|\Omega^{(n)}_\mathrm{CP2K}-\Omega^{(n)}_\mathrm{aims}|$ of BSE excitation energies $\Omexc$ computed from CP2K and FHI-aims by solving Eq.~\eqref{eq-BSE_ABBA} with BSE@ev$GW_0$@PBE.
    The mean absolute error~\eqref{eq-MAE_CP2K_aims} over the ten lowest excitation energies across all molecules is only 2.7 meV.
    For a benchmark on the impact of parameters of $GW$ calculations on BSE excitation energies, we refer to App.~\ref{app-thiels_set}. 
    }
    \label{fig-thiels_set_128+500}
\end{figure}

One focus of this work is to study the optical absorption spectra of finite nanostructures from $GW$-BSE.
Optical absorption in the linear-response regime is described by the dynamical dipole polarizability tensor $\alpha_{\mu\mu'}(\omega)$, which relates the induced electronic dipole polarization of the nanostructure $P_\mu(\omega)$ to an incident electric field $E_{\mu'}(\omega)$ of frequency $\omega$~\cite{Ullrich2011}:
\begin{align}
    P_\mu(\omega) = \sum_{\mu'} \alpha_{\mu\mu'}(\omega)\, E_{\mu'}(\omega), \quad  \mu,\mu'\in\{x,y,z\}.
\end{align}
In the BSE framework, $\alpha_{\mu\mu'}(\omega)$ can be computed from the solution $(\Omexc,\bX^{(n)},\bY^{(n)})$ of Eq.~\eqref{eq-BSE_ABBA} for $\alpha^\text{(S)}=2$~\cite{Ullrich2011, Jacquemin2016, Bruneval2016, McLACHLAN1964} (a derivation is also shown in App.~\ref{app-derivation_dyn_dip_pol_and_abba_eq}):
\begin{align}
    \alpha_{\mu\mu'}(\omega) 
    = - \sum_{n>0} \frac{2 \,\Omega^{(n)}\, d^{(n)}_{\mu} d^{(n)}_{\mu'}}{(\omega+i\eta)^2-\left(\Omega^{(n)}\right)^2} \, ,
    \label{eq-BSE_polarizatbility}
\end{align}
where the transition dipole moments of the Singlet solution are given by~\cite{Ljungberg2015, Bruneval2016, Jacquemin2016, Liu2020, McLACHLAN1964}
\begin{align}
    d^{(n)}_{\mu} = \sqrt{2} \sum_{i,a} \mu_{ia} \left(X_{ia}^{(n)} + Y_{ia}^{(n)}\right) \, ,
    \label{eq-BSE_trans_mom}
\end{align}
with the dipole operator in the KS orbital basis 
\begin{align}
    \mu_{ia} = \int\!d^3r \, \psi_i(\br) \, \br \,\psi_a(\br) \, . \label{eq-dipole_KS}
\end{align}
For the study of randomly oriented molecules in gases and liquids,    the isotropic average $\bar\alpha(\omega)$, the oscillator strengths $f^{(n)}$ and the photoabsorption cross section $\sigma_{\mu\mu'}(\omega)$ are typically used, which we discuss in App.~\ref{app-key_quantitities_for_opt_exc}.
In periodic crystals, the optical absorption is typically computed from the imaginary part of the (macroscopic) dielectric function $\epsilon (\omega) $~\cite{Adler1962, Wiser1963, Onida2002, Rohlfing2000} related to the current-current correlation function~\cite{Mahan2000, Rohlfing2000}.
In this work, we employ $\alpha_{\mu\mu'}(\omega)$ for computing optical absorption of  finite nanostructures, because of its structural similarity and relation to the dielectric function $\epsilon(\omega)$~\cite{Wiser1963, Strinati1988, Berger2005}, which is extracted from experimental data (cf. Sec.~\ref{sec-spectra_of_nanographenes_theory}).
Equivalently, one could include a factor of $\omega$ in the compared expressions, which would then be the absorption coefficient $\kappa_\text{abs} \propto \omega \epsilon(\omega)$~\cite{Strinati1988}, and the photoabsorption cross section $\sigma_{\mu\mu'}(\omega)\propto \omega \alpha_{\mu\mu'}(\omega)$.
To demonstrate the numerical precision of our \textit{GW}-BSE implementation in CP2K, we benchmark the excitation energies $\Omega^{(n)}$ from Eq.~\eqref{eq-BSE_ABBA} on Thiel's set~\cite{Schreiber2008}, a standard  set of organic molecules.
As reference implementation, we employ the $GW$-BSE implementation in FHI-aims~\cite{Blum2009, Ren2012, Liu2020, Abbott2025}.
We do not enforce identical quasiparticle energies as starting points, but both implementations are run from scratch as BSE@ev$GW_0$@PBE calculation with the aug-cc-pVDZ~\cite{Dunning1989,Kendall1992} basis set.
The computational details are given in App.~\ref{app-computational_details}. 
In Fig.~\ref{fig-thiels_set_128+500}, we report the absolute error~$|\Omega^{(n)}_\mathrm{CP2K}-\Omega^{(n)}_\text{FHI-aims}|$ for each molecule of Thiel's set and the first ten excitation levels $n = 1,\ldots, N_\mathrm{exc}$,  $N_\mathrm{exc}=10$, between  CP2K and FHI-aims.
The presented mean absolute error (MAE) is computed as
\begin{align}
    \mathrm{MAE} = \frac{1}{28\,N_\mathrm{exc}} \sum_{n=1}^{N_\mathrm{exc}} \sum_{M=1}^{28} |\Omega^{(n)}_{\mathrm{CP2K},M}-\Omega^{(n)}_{\text{FHI-aims},M}| \,, \label{eq-MAE_CP2K_aims}
\end{align}
where the index $M$ runs over the 28 molecules in the set and $\Omega^{(n)}_\mathrm{CP2K}$ and $\Omega^{(n)}_\mathrm{aims}$ are the extracted excitation energies from our implementation and the FHI-aims implementation~\cite{Liu2020}, respectively.
We observe a small MAE of 2.7~meV in Fig.~\ref{fig-thiels_set_128+500} with single outliers leading to a maximum error of 30~meV.
In general, the observed deviations between CP2K and FHI-aims can be traced back to the quasiparticle energies $\varepsilon_n^{\mathrm{ev}GW_0}$ around the HOMO-LUMO gap.
\section{Absorption spectrum of nanographenes from BSE and comparison to experiment}\label{sec-spectra_of_nanographenes_theory}
\begin{figure}
    \centering
    \includegraphics[width=\linewidth]{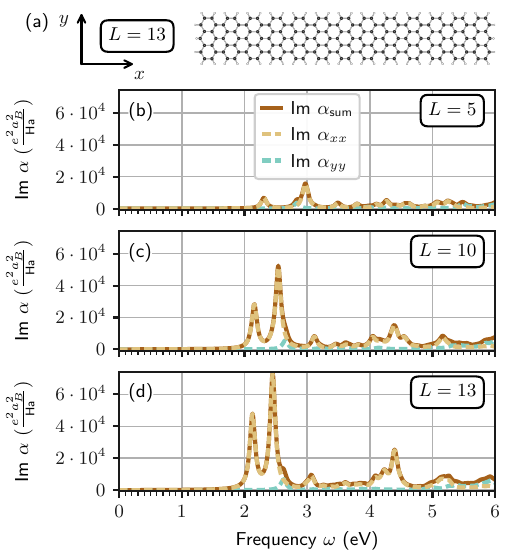}
    \caption{
    Optical absorption spectrum from the imaginary part of the dynamical dipole polarizability $\alpha_{\mu\mu'}(\omega)$~\eqref{eq-BSE_polarizatbility} ($\eta=0.05$~eV) for finite nanographene flakes with increasing number of repeating units $L$ from BSE@ev$GW_0$@PBE.
    (a) Geometry of a nanographene with $L=13$ anthracene units in $x$-direction\reviewin{, visualized using Ref.~\cite{Momma2011}.}
    (b-d) Components of the dynamical dipole polarizability and the sum, $ \alpha_\mathrm{sum} =    \alpha_{xx}+\alpha_{yy}+\alpha_{zz} $ as function of the light frequency.
    The out-of-plane component $\alpha_{zz}$ is not shown  as its 
     absolute value is below $\text{Im}\ \alpha_{zz} = 156~(e^2 a_B^2)/\mathrm{Ha}$.  
        Computational details are described in App.~\ref{app-computational_details}.
    }
    \label{fig-nanographene_geo_and_spectra}
\end{figure}

We now focus on optical excitations of rectangular nanographenes with seven carbon atoms in zigzag direction and an increasing number of repeating anthracene units, $L$, in armchair direction, as displayed in Fig.~\ref{fig-nanographene_geo_and_spectra}(a).
\reviewout{
Experimentally, the investigation of excitation spectra of such nanographenes is enabled by synthesizing them on noble metal surfaces~\cite{
}.
}

Since finite nanographenes feature spin-polarized zig-zag edge states when passivated by single hydrogen atoms~\cite{Wilhelm2018}, we introduce a second H atom at the outermost central C atoms along the $x$-direction, suppressing the edge states.
We compute optical properties from ev$GW_0$+BSE employing Eq.~\eqref{eq-BSE_ABBA}  when increasing the repeating units in armchair direction.
In Fig.~\ref{fig-nanographene_geo_and_spectra}, we show the imaginary part of the dynamical dipole polarizability tensor $\alpha_{\mu \mu'}$ from Eq.~\eqref{eq-BSE_polarizatbility} for three different number of repeating units $L=5,10,13$.
For all lengths $L$ in Fig.~\ref{fig-nanographene_geo_and_spectra}, we observe that the sum $\alpha_\mathrm{sum}(\omega)$ is dominated by the longitudinal $\alpha_{xx}(\omega)$ component.
While the $\alpha_{xx}(\omega)$ component shows a strong dependence on the system length $L$, we can confirm from Fig.~\ref{fig-nanographene_geo_and_spectra}(b-d) that the optical response in the transverse direction $\alpha_{yy}(\omega)$ remains largely unaffected by the length $L$ in $x$-direction, as it has also been observed in Ref.~\cite{Cocchi2012} from  semiempirical calculations.

\begin{figure}
    \centering
    \includegraphics[width=\linewidth]{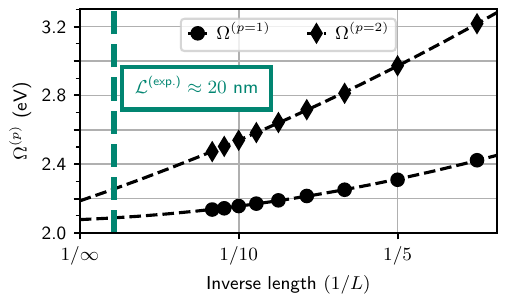}
    \caption{
    Excitation frequencies $\Omega^{(p=1,2)}$ of the first two dominant peaks ($p=1,2$) in $\mathrm{Im\ }{\alpha_{xx}}^{(L)}(\omega)$ for $L\in[4,13]$ (see example spectra in Fig.~\ref{fig-nanographene_geo_and_spectra}).
    We fit the the obtained peak frequencies by $\Omega^{(p)}(L) = a^{(p)} + b^{(p)}/L + c^{(p)}/L^2$ (dashed lines), from which we obtain $\li\Omega^{(p=1)} = 2.08$~eV and $\li \Omega^{(p=2)} = 2.19$~eV. 
    The obtained parameters of the fits for $p=1,2$ are reported in Table~\ref{tab-fit_parameters_peak_frequencies}.  
    The vertical dashed line in green indicates the average length of the experimental nanographene $\mathcal{L}^\text{(exp.)}\approx20$~nm~\cite{Denk2014}. 
    The fit yields $\Omega^{(p=1)}(\mathcal{L}^\text{(exp.)})=2.09$~eV  and $\Omega^{(p=2)}(\mathcal{L}^\text{(exp.)})=2.26$~eV .
    }
    \label{fig-fit_convergence_peak_frequency}
\end{figure}

\begin{figure}
    \centering
    \includegraphics[width=\linewidth]{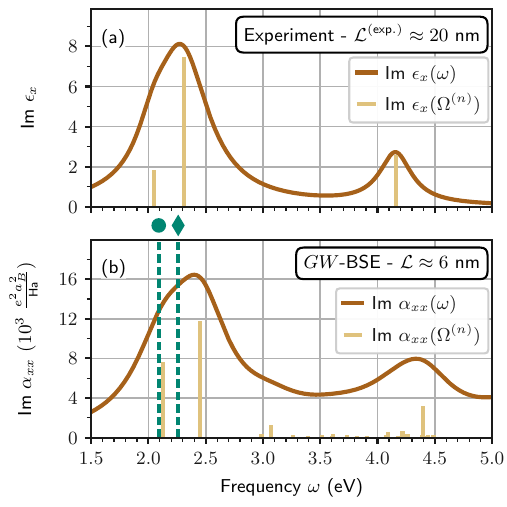}
    \caption{
    Optical absorption spectrum (a) reproduced from experimental data~\cite{Denk2014} via $\text{Im}\ \epsilon_x(\omega)$ and (b) calculated using $GW$-BSE (BSE@ev$GW_0$) for $L=13$ via $\text{Im}\ \alpha_{xx}(\omega)$, which are related by a frequency-independent factor~$C$ [Eq.~\eqref{eq-relation_prop_epsilon_alpha}]. 
    For the $GW$-BSE calculation, we employ a broadening of $\eta=0.3$~eV.
    The individual peak heights at the resonance energies~$\Omega^{(n)}$\reviewin{, represented by the beige vertical lines,} are computed from Eq.~\eqref{eq-absorption_peaks_from_exp}/\eqref{eq-absorption_peaks_from_theory_as_epsilon} in App.~\ref{app-diel_func_from_pol}.  
    Green vertical dashed lines denote the extrapolated peak frequencies $\Omega^{(p=1)}(\mathcal{L}^\text{(exp.)})=2.09$~eV and $\Omega^{(p=2)}(\mathcal{L}^\text{(exp.)})=2.26$~eV  from Fig.~\ref{fig-fit_convergence_peak_frequency}.
    }
    \label{fig-abs_spectra_BSE_vs_exp}
\end{figure}

The nanographene studied in Ref.~\cite{Denk2014} is about 20\,nm long, which corresponds to $L=46$ and is beyond the system sizes accessible to our current BSE implementation.
To compare with the experimental optical absorption spectrum of Ref.~\cite{Denk2014}, we extract the frequencies of the two lowest longitudinal peaks, $\Omega^{(p)}$ with $p=1,2$, from $\mathrm{Im}\,\alpha_{xx}^{(L)}(\omega)$ for $L=4,\dots,13$ and extrapolate the resulting $\Omega^{(p)}(L)$ to the experimental geometry.
The data points used for this procedure are shown in Fig.~\ref{fig-fit_convergence_peak_frequency}.
We omit $L\le 3$ because the corresponding spectra do not yet exhibit two well-separated low-energy peaks that would yield reliable peak positions.
For the extrapolation, we apply a quadratic fit~$\Omega^{(p)}(L) = a^{(p)} + b^{(p)}/L + c^{(p)}/L^2$  motivated by the free-electron model in a one-dimensional box with periodic potential, as it has been already applied to other $\pi$-electron systems~\cite{deMelo1999, Torras2012}.
\begin{table}[t!]
  \caption{Fit parameters of the quadratic fit $\Omega^{(p)}(L) = a + b/L + c/L^2$ in Fig.~\ref{fig-fit_convergence_peak_frequency} and extrapolated excitation energies and experimental measurement (cf. Fig.~\ref{fig-abs_spectra_BSE_vs_exp}). All quantities are given in eV.}
    \centering
    \begin{tabular}{lccccc}
    \hline\hline
        &  \;\;\;\;\;$a^{(p)}$\;\; &\;\; $b^{(p)}$\;\;  & \;\;$c^{(p)}$ \;\;& \;\;$\Omega^{(p)}(\mathcal{L}^\text{(exp.)})$\;\;& \;\;$\Omega^{(p)}  $ (exp.) \\ \hline
        $p=1$ & 2.08 & 0.39 & 3.96       &   2.09 & 2.05  \\
        $p=2$ & 2.19 & 3.13 & 3.97       &   2.26 & 2.31\\
        \hline\hline
    \end{tabular}
    \label{tab-fit_parameters_peak_frequencies}
\end{table}
Taking into account the finite size of the experimental geometries (green vertical line in Fig.~\ref{fig-fit_convergence_peak_frequency}), which they estimate to be $\mathcal{L}^\text{(exp.)}\approx20$~nm (cf. supporting information of Ref.~\cite{Denk2014}), we obtain BSE excitation energies of $\Omega^{(p=1)}(\mathcal{L}^\text{(exp.)})=2.09$~eV  and $\Omega^{(p=2)}(\mathcal{L}^\text{(exp.)})=2.26$~eV from the extrapolation.
These values are in excellent agreement with  the experimental excitation energies~\cite{Denk2014}  $\Omega^{(p=1)}_{\mathrm{exp.}} = 2.05$~eV and $\Omega^{(p=2)}_{\mathrm{exp.}} = 2.31$~eV, yet another demonstration of the success of the BSE formalism.
Beyond the agreement of the peak positions, we compare the peak heights of the $GW$-BSE and experimental optical absorption spectra.
Ref.~\cite{Denk2014} obtains the dielectric function~$\epsilon_x(\omega)$ of the nanographene (20 nm length) experimentally via reflectance difference measurements, whose imaginary part is shown in Fig.~\ref{fig-abs_spectra_BSE_vs_exp}(a) (see details in App.~\ref{app-diel_func_from_pol}). 
In the limit of a infinitely long nanographene, the dielectric function is related to the dynamical dipole polarizability via a frequency-independent factor~\cite{Berger2005}:
\begin{align}
    \text{Im}\ \epsilon_x(\omega) = C \ \text{Im}\ \alpha_{xx}(\omega)\,,
    \label{eq-relation_prop_epsilon_alpha}
\end{align}
which has the same form as the \textit{Clausius-Mossotti} relation when neglecting local field effects~\cite{Wiser1963}.
Hence, we plot the imaginary part of the dynamical dipole polarizability $\text{Im}\ \alpha_{xx}(\omega)$ in Fig.~\ref{fig-abs_spectra_BSE_vs_exp}(b) for the nanographene with $L=13$ (6\,nm length).
Comparing the experimental and BSE absorption spectrum in Fig.~\ref{fig-abs_spectra_BSE_vs_exp}(a) and (b), we observe a nice match of the overall lineshape.
The observable difference between experiment and BSE spectrum, particularly the resonance frequencies $\Omega^{(n)}$, may originate from the different lengths in the experimental and $GW$-BSE geometries (20 nm versus 6 nm).
When extrapolating  BSE excitation energies to 20 nm length, the agreement between BSE and experiment is excellent, cf.~Table~\ref{tab-fit_parameters_peak_frequencies}.

\section{Spatial descriptors of electronic excitations within the BSE}\label{sec-theory_of_descriptors}
Beyond the optical response,  excited states  have an associated spatial structure, which can be of interest to better understand oscillator strengths or transition rates in excited-state dynamics.
To enable a quantitative analysis of these spatial properties, we discuss expectation values (\textit{descriptors}) associated with particular spatial properties of an excitation in this section, as proposed in Refs.~\cite{Bappler2014, Plasser2015, Mewes2015, Mewes2018}.
The starting point is the wave function of the electronic excitation~\cite{Mewes2018},
\begin{align}
    \Psi\exc(\bre, \brh) = \int\!\text{d}^3r_2\dots\text{d}^3r_N \Phi_0(\brh,\br_2,\dots) \Phi_n(\bre,\br_2,\dots) \, , \label{eq-exc_wave_function_from_manybody_wavefunctions}
\end{align}
where $\Phi_0$ and $\Phi_n$ are the ground and an excited state wave function of the full many-body Hamiltonian.
Within $GW$-BSE, Eq.~\eqref{eq-exc_wave_function_from_manybody_wavefunctions} can be expressed in the product space of occupied and unoccupied KS orbitals as~\cite{Mewes2015, Rohlfing2000, Jacquemin2016}
\begin{align}
    \Psi\exc (\bre, \brh) 
    = \sum_{i,a} \Xian \psi_a(\bre) \psi_i(\brh) + \Yian \psi_i(\bre) \psi_a(\brh) \, . \label{eq-exc_wavefunction}
\end{align}
The physics behind Eq.~\eqref{eq-exc_wavefunction} can be understood in terms of electrons and holes:
Concerning the first part, an electron is excited to a (formerly) unoccupied orbital $\psi_a(\bre)$ and leaves a hole behind at the occupied orbital $\psi_i(\brh)$, where the excitation $n$ can mix different KS orbitals as described by the transition amplitudes $\Xian$.
Vice versa, the amplitudes $\Yian$ describe the reverse process, i.e.~deexcitations~\cite{Dreuw2005}.
The expectation value associated with excitation~\eqref{eq-exc_wavefunction} for a generic operator $\hat{O}$ then reads
\begin{align}
\langle \hat{O} \rangle \exc
=
\frac{ 
 \langle \Psi\exc | \hat{O} | \Psi\exc\rangle 
}{
 \langle \Psi\exc | \Psi\exc\rangle 
}
\, . \label{eq-expectation_value_excitation}
\end{align}
From here on, we drop the excitation index $n$ to simplify the notation of the expectation values.
Specifically, we focus on the size of the excitation~\cite{Mewes2018}
\begin{align}
    d_\text{exc} = \sqrt{\langle | \bre - \brh|^2\rangle_\text{exc}} \, , \label{eq-exc_size}
\end{align}
which can be divided into the following contributions:
\begin{align}
    d_\text{exc} = \sqrt{d_{e\rightarrow h}^2 + \sigma_h^2 + \sigma_e^2 - 2 \sigma_h \sigma_e R_{eh} } \, . \label{eq-exc_size_contributions}
\end{align}
The first contribution is the distance of electron and hole 
\begin{align}
    d_{e\rightarrow h} = |\langle \bre - \brh\rangle_\text{exc}| \, , \label{eq-distance_eh}
\end{align}
which naturally quantifies the charge-transfer character of an excitation.
The electron-hole separation $d_{e\rightarrow h}$ is large for a so-called a charge-transfer excitation~\cite{Mewes2015, Mewes2018} where the electron at $\langle \bre\rangle_\text{exc}$ is located on a different part of the molecule~\cite{Plasser2012,Blase2011, Blase2018, Loos2021, Knysh2023} or of a van-der-Waals crystal~\cite{Deilmann2018, Naik2022, Chan2024} than the hole at $\langle \brh\rangle_\text{exc}$.
Further contributions to  Eq.~\eqref{eq-exc_size_contributions}   are the size of the electron
\begin{align}
    \sigma_e = \sqrt{\langle (\bre - \langle \bre \rangle_\text{exc})^2 \rangle_\text{exc}} \, , \label{eq-size_electron}
\end{align}
 analogously the size of the hole, $\sigma_h$, and the electron-hole correlation coefficient $R_{eh}$
\begin{align}
    R_{eh} = \frac{1}{\sigma_e \sigma_h} \left( \langle \mathbf{r}_h \cdot \mathbf{r}_e \rangle_\mathrm{exc}
- \langle \mathbf{r}_h \rangle_\mathrm{exc} \cdot \langle \mathbf{r}_e \rangle_\mathrm{exc} \right) \, , \label{eq-correlation_coeff}
\end{align}
If $R_{eh}$ is nonzero, it allows us to distinguish between bound electron-hole-pairs ($R_{eh} > 0$) and pairs, which are avoiding each other dynamically ($R_{eh}<0$).
Consequently, $R_{eh}=0$ indicates uncorrelated electron and holes.
By the Cauchy-Schwarz inequality, we find that $R_{eh}\in[-1,1]$, i.e.~maximal anti-/correlation is indicated by --1/+1, respectively.
To describe the strongly anisotropic nanographenes, we additionally define \textit{directional exciton descriptors}, where the coordinates of electron and hole are projected into the cartesian directions.
This leads, exemplarily for the $x$-direction, to the size of the electron
\begin{align}
    \sigma_e^{(x)} = \sqrt{\langle (x_e - \langle x_e \rangle_\text{exc})^2 \rangle_\text{exc}} \, , \label{eq-longitudinal_electron_size}
\end{align}
and by analogy for $\sigma_h^{(x)}$, the (directional) electron-hole separation
\begin{align}
    d_{e\rightarrow h}^{(x)} = |\langle x_e - x_h \rangle_\text{exc}| \, , 
\end{align}
as well as the size of the excitation in $x$-direction
\begin{align}
    d_\text{exc}^{(x)} = \sqrt{\langle | x_e - x_h|^2\rangle_\text{exc}} \, . \label{eq-longitudinal_excitation_size}
\end{align}
We define the  correlation coefficient in $x$-direction as
\begin{align}
    R_{eh}^{(x)} 
    = \frac{1}{\sigma_e^{(x)} \sigma_h^{(x)}} \left( \langle x_h \cdot x_e \rangle_\mathrm{exc}
- \langle x_h \rangle_\mathrm{exc} \cdot \langle x_e \rangle_\mathrm{exc} \right) \, . \label{eq-correlation_coeff_directional}
\end{align}
We implement all descriptors based on multipole moments in the KS orbital basis following Ref.~\cite{Mewes2018}\footnote{
Exemplarily, the individual expectation values for the average position of hole $\brh$ and electron $\bre$ in an excited state $n$ are computed from their cartesian components ($\mu\in\{x,y,z\}$)
\begin{align*}
    \langle r_e^{(\mu)} \rangle_\mathrm{exc} 
    &= 
\frac{ 
 1
}{
 C_n 
}
    \left(
    \sum_{a,i,j}
    \Xian \Xjan \mu_{ij} 
    + \sum_{a,b,i} \Yian \Yibn \mu_{ab} \right) 
    \\
    \langle  r_h^{(\mu)} \rangle_\mathrm{exc} 
    &= 
\frac{ 
 1
}{
 C_n
}
    \left(
     \sum_{a,b,i} \Xian \Xibn \mu_{ab} 
    + \sum_{a,i,j} \Yian \Yjan \mu_{ij} \right)
    \, .
\end{align*}
Here, we have introduced the dipole moments $\mu_{pq}=\langle \psi_p|\hat{\mu}| \psi_q \rangle$ in the KS orbitals $\psi_p(\br)$, which are computed in the length gauge (cf. Eq.~\eqref{eq-BSE_trans_mom}), as well as the normalization factor
\begin{align*}
    C_n 
    &=  \langle \Psi\exc | \Psi\exc\rangle 
    = \sum_{i,a} ( X_{ia}^{(n)} )^2 + ( Y_{ia}^{(n)} )^2\\
    &=  1 + 2 \sum_{i,a}( Y_{ia}^{(n)} )^2 \geq1 ,
\end{align*}
where we have used Eq.~\eqref{eq-normalization_XY}~\cite{Plasser2025}.
Hence, $C_n=1$ for all $n$ in the case of the TDA.
}.

\section{Size of excitations in nanographenes from BSE and TDDFT} \label{sec-spatial_descriptors}

\begin{figure}[t]
    \centering
    \includegraphics[width=\linewidth]{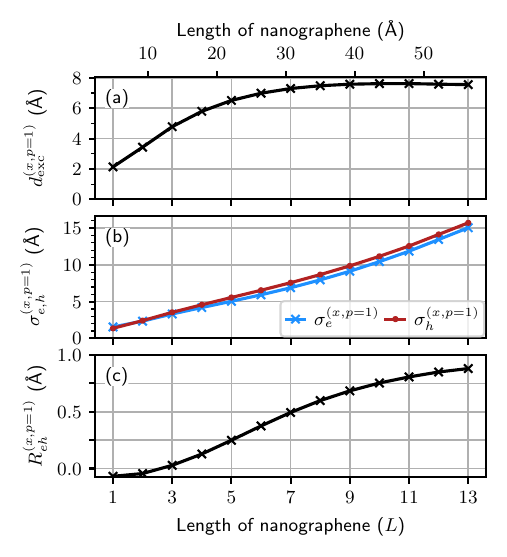}
    \caption{
    Spatial properties of the first bright peak $p=1$ in the optical absorption spectra along the $x$-direction computed from $GW$-BSE.
    (a) Longitudinal size of the excitation at the first peak $d_\mathrm{exc}^{(x,p=1)}$, Eq~\eqref{eq-longitudinal_excitation_size}, 
    (b) longitudinal sizes of electron $\sigma_e^{(x,p=1)}$ (blue) and hole $\sigma_h^{(x)}$ (red) from Eq.~\eqref{eq-longitudinal_electron_size}, 
    (c) longitudinal electron-hole correlation coefficient $R_{eh}^{(x,p=1)}$~\eqref{eq-correlation_coeff_directional} associated with the first peak $p=1$.
    }
    \label{fig-descriptors_bse}
\end{figure}

In this section, we analyze spatial characteristics of the electronic excitations in the nanographenes of Sec.~\ref{sec-spectra_of_nanographenes_theory}, focusing on the lowest-energy bright state ($p=1$) for each length $L$.
The size of the excitation $d_\mathrm{exc}^{(x,p=1)}$ is shown in Fig.~\ref{fig-descriptors_bse}(a) as a function of the nanographene length. 
For small $L\in[1,4]$, we observe a linear increase, which saturates for larger system sizes to $\sim$\,7.6~\AA\  for a length of 4~nm.
The plateau indicates that the size of the excitation is controlled by an intrinsic correlation length (exciton radius) that is smaller than the nanographene length, consistent with bound excitons in crystals.
In contrast, the individual spreads of electron and hole, $\sigma_e^{(x,1)}$ and $\sigma_h^{(x,1)}$, increase with $L$ [Fig.~\ref{fig-descriptors_bse}(b)], as expected for single-particle densities in a one-dimensional box, which extend over the system whereas the relative separation of electron and hole is bounded via the screened Coulomb interaction. 
The correlation coefficient in Fig.~\ref{fig-descriptors_bse}(c) confirms this view: it increases with $L$ and approaches one, indicating strongly correlated electron-hole motion for large $L$.

In order to assess to which extent a simpler framework can reproduce these excitonic characteristics, we next turn to TDDFT~\cite{Ullrich2011}, which offers a computationally  cheaper alternative to the \textit{GW}-BSE approach.
Beyond that, it is available in most electronic-structure codes, making it attractive for systematic studies of large nanographenes.
However, results from TDDFT computations depend strongly on the employed exchange–correlation functional, which implicitly determines the electron–hole interaction~\cite{Mewes2017}.
In semi-local exchange-correlation kernels in the generalized gradient approximation (GGA), this interaction is absent, while hybrid and range-separated hybrid functionals reintroduce parts of the bare Coulomb attraction through exact exchange.
For example, PBE0 includes 25\% of the unscreened interaction, whereas HSE06 restricts it to short range, effectively mimicking a fully screened Coulomb potential in the long range.
\reviewin{For metallic systems, TDDFT with (semi-)local functionals captures the features of plasmonic excitations~\cite{Quong1993, Gurtubay2005,Ullrich2011}.
This is commonly attributed to strong metallic screening, which suppresses  the interaction of the excited electrons and holes.
In gapped materials, however, (semi-)local TDDFT kernels typically miss the required long-range electron-hole attraction and therefore yield inaccurate excitation spectra (e.g., for Si)~\cite{Onida2002,Botti2004}.
Within TDDFT, a reliable description then requires a suitably nonlocal exchange-correlation kernel, for instance based on (range-separated) hybrid functionals or other long-range corrected kernels.

}

\begin{figure}
    \centering
    \includegraphics[width=\linewidth]{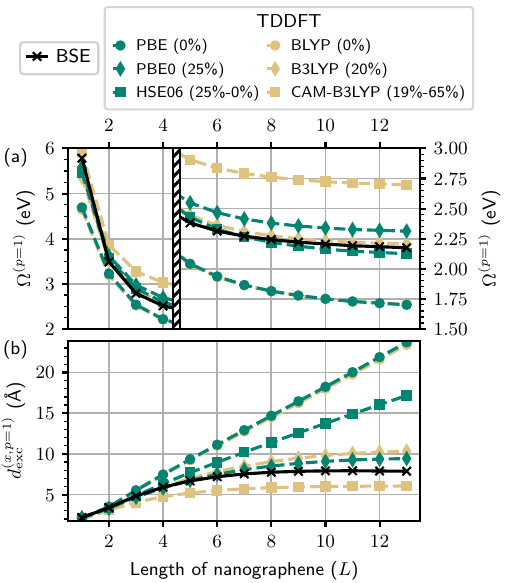}
    \caption{
    Excitation energy $\Omega^\text{(p=1)}$ and size of the excitation $d_\mathrm{exc}^{(x,\text{p=1})}$ computed from BSE@ev$GW_0$ and TDDFT (both in TDA; effect of TDA investigated in App.~\ref{app-bse_abba_vs_TDDFT_tda}) employing different exchange–correlation functionals (exact exchange fraction shown in the legend).
    (a) Peak frequency of the first prominent peak in the optical absorption spectrum $\Omega^\text{(p=1)}$.
    (b) Longitudinal size of the excitation for the first bright peak $d_\mathrm{exc}^{(x,\text{p=1})}$, Eq.~\eqref{eq-longitudinal_excitation_size}.
   We show excitation energies and sizes computed with PBE-based hybrids as a function of the exact-exchange fraction in App.~\ref{app-tddft_with_pbeh}.
    }
    \label{fig-tddft_gga_hyb_rs_hyb_vs_BSE}
\end{figure}

Figure~\ref{fig-tddft_gga_hyb_rs_hyb_vs_BSE} compares excitation energies and sizes of the excitation obtained \reviewin{in nanographenes} from TDDFT with those from \textit{GW}-BSE.
In panel (a), all methods show the expected redshift of the first bright excitation with increasing nanographene length, but the absolute values differ.
GGAs (PBE~\cite{Perdew1996}, BLYP~\cite{Lee1988,Becke1988}) substantially underestimate the excitation energies, while hybrids (PBE0, B3LYP~\cite{Lee1988,Becke1993,Stephens1994}, HSE06 ~\cite{Heyd2003,Krukau2006}) yield values closer to $GW$-BSE.
The exception is CAM-B3LYP~\cite{Yanai2004}, overestimating the BSE excitation energy by $\sim$\,0.5~eV, which we attribute to the large long-range exact-exchange fraction of 65\,\%, binding the exciton too strongly.
Panel~(b) shows the corresponding sizes of the excitation $d_\mathrm{exc}^{(x,\text{p=1})}$: GGAs predict overly delocalized excitations whose size grow linearly with $L$, which we attribute to the absent attractive Coulomb interaction between electron and hole.
Hybrid functionals confine the excitation more strongly, although HSE06 still shows an increase of the size of the excitation with nanographene length since the Coulomb interaction of electron and hole is only present in short range.
Overall, our TDDFT calculations reproduce qualitative trends of the BSE only when the functional contains sufficient long-range electron-hole attraction, but its quantitative accuracy remains limited by the absence or approximate treatment of screening.
\reviewin{Accordingly, manual tuning of functional parameters to mimic the screened long-range electron-hole attraction should also enable better quantitative prediction of excitation energies, oscillator strengths and spatial descriptors.
We discuss such a tuning for the exact-exchange fraction in PBE-based hybrid functionals in App.~\ref{app-tddft_with_pbeh}.
By analogy, functionals with more parameters might allow to match several properties simultaneously for a given material.
However, the loss of generality caused by this manual tuning for a given material strongly limits its applicability to novel materials.
}

\begin{figure}
    \centering
    \includegraphics[width=\linewidth]{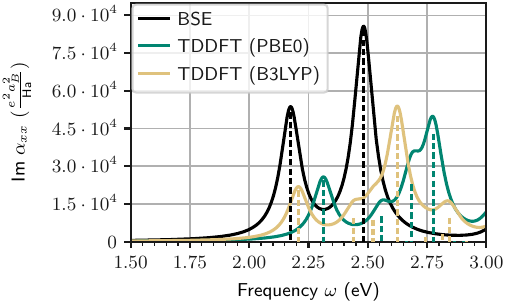}
    \caption{
    Optical absorption spectrum  for $GW$-BSE (black) and TDDFT with two different hybrid functionals PBE0 (green) and B3LYP (beige) for the nanographene with $L=13$ \reviewin{(all in TDA; effect of TDA investigated in App.~\ref{app-bse_abba_vs_TDDFT_tda})}, computed from Eq.~\eqref{eq-BSE_polarizatbility} \reviewin{employing $\eta=0.05$~eV}.
    Dashed lines  denote the contribution of individual excitations $n$ to $ \mathrm{Im}\ \alpha_{xx}(\omega=\Omega^{(n)})$ from Eq.~\eqref{eq-BSE_polarizatbility} (cf. Eq.~\eqref{eq-absorption_peaks_from_theory_as_epsilon}).
    }
    \label{fig-tddft_vs_bse_spectra_comparison}
\end{figure}
Fig.~\ref{fig-tddft_gga_hyb_rs_hyb_vs_BSE} establishes which hybrid functionals yield a reasonable excitation energy and corresponding size of the lowest bright excitation; however, their ability to reproduce the overall spectral shape has not yet been assessed.
To assess how well TDDFT reproduces the full optical spectra beyond the lowest excitation, we now compare the absorption spectra from TDDFT and \textit{GW}-BSE in Fig.~\ref{fig-tddft_vs_bse_spectra_comparison}. 
For the nanographene with $L=13$, the TDDFT spectra obtained with PBE0 and B3LYP differ significantly from the \textit{GW}-BSE reference (Fig.~\ref{fig-tddft_vs_bse_spectra_comparison}): the number and relative intensity of peaks are inconsistent, and additional side peaks appear both at lower and higher energies. 
Assuming \textit{GW}-BSE as the reference, these discrepancies indicate that the TDDFT excitation spectrum is qualitatively incorrect, which we attribute to the incorrect interaction of electron and hole in our TDDFT calculations. 
Recently advances for range-separated hybrid functionals~\cite{Camarasa-Gomez2023, Ohad2023, Camarasa-Gomez2024} might be able to mimic the screened interaction more accurately, but their application is beyond the scope of this work. 
In any case, the large experimental linewidths (cf. Fig.~\ref{fig-abs_spectra_BSE_vs_exp}) would likely wash out these differences even for standard functionals, such that the broadened TDDFT spectra could still appear compatible with experiment despite the incorrect underlying peak structure.

\section{Conclusion}
We have implemented and applied the \textit{GW}-BSE formalism in CP2K to rectangular nanographenes with a fixed width of seven carbon atoms along the zigzag direction and variable length along the armchair direction. 
The BSE@ev$GW_0$@PBE approach reproduces the measured absorption spectrum, and the two lowest bright excitations agree with experiment within 0.05~eV after finite-size extrapolation. 
The size of the excitation of the first bright state converges to about 7.6~\AA, establishing a bound exciton with a well-defined spatial extent.
TDDFT with PBE0 and B3LYP reproduces this size of the excitation and yields reasonable lowest excitation energies, but the spectra around the lowest bright peaks display additional sidebands and altered intensities that are absent in \textit{GW}-BSE. 
Thus, while selected descriptors can be matched within TDDFT, \textit{GW}-BSE provides a consistent description of both spectral and spatial properties for nanographenes.

\section*{Data and Code availability}
Inputs and outputs of all calculations reported in this work are available in a GitHub~\cite{GithubData} and a Zenodo~\cite{ZenodoData} repository, as well as via the University of Regensburg library~\cite{RegensburgData}.
The $GW$+BSE algorithm developed in this work is available in the open-source package CP2K~\cite{CP2K,Kuhne2020, Iannuzzi2026}.
\section*{Acknowledgements}
We thank Deborah Prezzi, Carlo Pignedoli, Anna Hehn, Wolfgang Hogger, Adrian Seith, Torsten Weber and in particular Štěpán Marek for helpful discussions.
The German Research Foundation (Deutsche Forschungsgemeinschaft, DFG) is acknowledged for funding via the Emmy Noether Programme (project number 503985532) and RTG 2905 (project number 502572516).
The authors gratefully acknowledge the computing time made available to them on the high-performance computers Noctua 2~\cite{Bauer2024} \reviewin{and Otus~\cite{Ehtesabi2025}} at the NHR Center Paderborn Center for Parallel Computing (PC2). This center is jointly supported by the Federal Ministry of Research, Technology and Space and the state governments participating in the National High-Performance Computing (NHR) joint funding program (\url{www.nhr-verein.de/en/our-partners}).
\reviewin{The support of Robert Schade and Xin Wu from PC2 with compiling and optimizing the CP2K build is gratefully acknowledged.}

\appendix
\section{Hedin's equations and commonly applied simplifications}\label{app-gw_schemes}
\reviewin{
In Sec.~\ref{sec-theory}, we have introduced the $GW$ approximation to the self-energy in Eq.~\eqref{eq-GW}, which reads in simplified notation~\cite{Hedin1965,Golze2019}
\begin{align}
    \Sigma^{GW}(\mathbf{1},\mathbf{2}) = 
    i G(\mathbf{1},\mathbf{2}) W(\mathbf{1}^+,\mathbf{2}) \, , \label{eq-GW_app_A}
\end{align}
where the numbers denote combined space-spin-time coordinates, e.g., $\mathbf{1}=(\br_1,s_1,t_1)$ and $\mathbf{1}^+$ denotes an infinitely small time-shift on $t_1$ to a later time. 
More generally, the full set of Hedin's equations reads~\cite{Golze2019}
\begin{align}
\Sigma(\mathbf{1},\mathbf{2}) &= i\int\! d(\mathbf{3}\mathbf{4})\, G(\mathbf{1},\mathbf{4})\,W(\mathbf{1}^+,\mathbf{3})\,\Gamma(\mathbf{4},\mathbf{2},\mathbf{3})\,,\label{eq-hedins_self_energy}
\\
G(\mathbf{1},\mathbf{2}) &= G_\mathbf{0}(\mathbf{1},\mathbf{2}) + \int\! d(\mathbf{3}\mathbf{4})G_\mathbf{0}(\mathbf{1},\mathbf{3})\Sigma(\mathbf{3},\mathbf{4})G(\mathbf{4},\mathbf{2})\,,\label{eq-GfHedin}
\\
W(\mathbf{1},\mathbf{2}) &=v(\mathbf{1},\mathbf{2}) + \int\! d(\mathbf{3}\mathbf{4}) \chi_0(\mathbf{3},\mathbf{4}) v(\mathbf{1},\mathbf{3}) W(\mathbf{4},\mathbf{2})\,, \label{eq-WHedin}
\\
\chi_0(\mathbf{1},\mathbf{2}) &=-i \int\! d(\mathbf{3}\mathbf{4}) G(\mathbf{4},\mathbf{2})G(\mathbf{2},\mathbf{3})\Gamma(\mathbf{3},\mathbf{4},\mathbf{1})\,,\label{eq-polHedin}
\\
\Gamma(\mathbf{1},\mathbf{2},\mathbf{3}) &=\delta(\mathbf{1},\mathbf{2})\delta(\mathbf{1},\mathbf{3})\nonumber
\\
&+ \int\! d(\mathbf{4}\mathbf{5}\mathbf{6}\mathbf{7}) \frac{\partial\Sigma(\mathbf{1},\mathbf{2})}{\partial G(\mathbf{4},\mathbf{5})}G(\mathbf{4},\mathbf{6})G(\mathbf{7},\mathbf{5})\Gamma(\mathbf{6},\mathbf{7},\mathbf{3})\,,
\end{align}
where the $GW$ approximation, Eq.~\eqref{eq-GW_app_A}, follows from Eq.~\eqref{eq-hedins_self_energy} when assuming $\Gamma(\mathbf{1},\mathbf{2},\mathbf{3})=\delta(\mathbf{1},\mathbf{2}) \delta(\mathbf{1},\mathbf{3})$.
The resulting $GW$ equations can be solved fully self-consistently (\textit{fully self-consistent $GW$}, sc$GW$).
However, it has been observed that sc$GW$ exhibits worse accuracy than other (partially) self-consistent schemes, related to an error cancellation observed for non/partial self-consistency and missing vertex corrections in $GW$~\cite{Onida2002, Golze2017, Martin2016}.
As a result, single-shot and partially self-consistent $GW$ schemes emerged.
The simplest scheme is single-shot $G_0W_0$, where both the Green's function~$G_0$ as well as the screened Coulomb interaction~$W_0$ are computed only once from the initial KS states and energies.
Due to missing self-consistency, $G_0W_0$ shows a strong starting point dependence on the exchange-correlation (xc) functional used in the preceding KS-DFT calculation~\cite{Blase2011a, Jacquemin2015, Bruneval2015, Caruso2016, Kaplan2016, Jacquemin2017a, Gui2018, Golze2019, Knysh2024}.
By partially introducing  self-consistency in $G$ via updates of the eigenvalues, but keeping the screened Coulomb interaction fixed as $W_0$ (ev$GW_0$), the starting point dependence is reduced and it has been shown that bandgaps of crystals computed from ev$GW_0$ are in best agreement to experiments  among the $GW$ schemes~\cite{Shishkin2007, Jiang2016, Grumet2018, Golze2019}.
For small molecules, $GW$ schemes are benchmarked against energies of the highest occupied and lowest unoccupied molecular orbital (HOMO, LUMO) computed from high-level wavefunction theory.
For molecules, it has been found that updating the screened interaction $W$ (ev$GW$) is on average slightly more favorable than keeping it fixed as $W_0$~\cite{Jacquemin2015,Gui2018,Kaplan2016}.
While nanographenes have a size between the limit of small molecules and extended crystals, their excitations have a spatial extent comparable to those in crystals.
Consequently, we use the ev$GW_0$ scheme in this work.
It has been found that partially self-consistent approaches (ev$GW_0$, ev$GW$) still show a starting point dependence, which has been observed to be around 50-250~meV~\cite{Knysh2024, Jacquemin2015, Gui2018, Forster2022} in molecular systems or even more than 500~meV for halide perovskites~\cite{Leppert2019}, since KS orbitals depend on the employed xc functional.
%
This remaining starting point dependence can be eliminated by updating the quasiparticle wavefunctions self-consistently  (qs$GW_0$, qs$GW$).
This approach has been successfully employed, e.g. by Ref.~\cite{Chen2014}, but that is beyond the scope of this work.

\begin{table*}[tb!]
 \setlength{\tabcolsep}{5pt}
 \def\arraystretch{1.5}
    \centering
        \caption{ 
        \reviewin{
        HOMO-LUMO gaps and first excitation energies (all in eV)  for the molecules ethene, benzene, and the $L=1$ nanographene, employing  PBE~\cite{Perdew1996} and PBE0~\cite{Adamo1999} as starting point.
        The excited state methods all employ the Tamm Dancoff approximation (TDA) for comparability.
        }
        }
    \begin{tabular}{l | cc||cc|cc|cc}
     &\multicolumn{2}{c||}{Eq.~\eqref{eq-contributions_ABBA_app_a}}&\multicolumn{2}{c|}{Ethene}&\multicolumn{2}{c|}{Benzene}&\multicolumn{2}{c}{Nanographene with $L=1$}
    \\
    &$\varepsilon_p$&$F_{pq,rs}$& PBE  & PBE0 & PBE  & PBE0 & PBE  & PBE0
    \\
\hline
\hline
$\Delta\varepsilon^\text{KS}$& - & - & 
5.64 & 7.77 & 5.11 & 6.99 & 4.14 & 5.92 
\\
$\Delta\varepsilon^{\text{ev}GW_0}$& - & - & 
11.19 & 11.14 & 9.66 & 9.64 & 8.65 & 8.71
\\
$\Delta\varepsilon^{\text{ev}GW}$& - & - & 
11.47 & 11.30 &  9.98 & 9.80 & 9.01 & 8.88
\\
$\Omega^{(n=1)}_\text{BSE@ev$GW_0$}$ 
&ev$GW_0$& $-W_{pq,rs}$& 6.24 & 6.44 & 4.84 & 4.98 & 4.62 & 4.74 
 \\
$\Omega^{(n=1)}_\text{TDDFT}$ 
&KS& $f^\text{xc}_{pq,rs}$& 6.48 & 6.87 & 5.24 & 5.48 & 4.35 & 4.97
 \\
$\Omega^{(n=1)}_{\text{RPA@KS}}$ 
&KS&0& 6.61 & 8.12 & 5.41 & 7.30 & 4.40 & 6.19
 \\
$\Omega^{(n=1)}_{\text{RPA@ev}GW}$ 
&ev$GW$&0 &
11.59 & 11.40 & 10.03 & 9.84 & 9.04 & 8.90
 \\
    \end{tabular}
    \label{tab-excitation_energies_TDDFT_RPA_BSE}
\end{table*}

Table~\ref{tab-excitation_energies_TDDFT_RPA_BSE} summarizes the starting-point dependence of the $GW$ HOMO-LUMO gaps and BSE@ev$GW_0$ excitation energies, using the DFT functionals PBE~\cite{Perdew1996} and PBE0~\cite{Adamo1999} as starting points for ethene, benzene and a nanographene consisting of a single unit cell ($L=1$; 9,10-dihydroanthracene).
In general, we confirm the starting point dependency between 50 and 250~meV, as given above. 
Moreover, the ordering of HOMO-LUMO gaps and excitation energies (ethene $>$ benzene $>$ nanographene) is identical for every starting point.
The good agreement of ev$GW_0$ bandgaps of crystals with experiments can be reasoned with in a simple picture:
We consider the effect of an external perturbation potential $V_\text{ext}(\br,t)$ on the electronic system (in linear response), given by the change of the electronic density~\cite{Onida2002, Ullrich2011}
\begin{align}
    \Delta n(\br,t) = \int\!\text{d}\br'\,\int\!\text{d}t'\, \chi(\br,t,\br',t')\,V_\text{ext}(\br,t) \,,
\end{align}
where the reducible polarizability (or density-density response function) $\chi$ is exact and includes the poles of the full many-body wavefunction.
$\chi$ follows from the irreducible polarizability $\chi_0$~\eqref{eq-polHedin} in Hedin's equations via
\begin{align}
    \chi(\mathbf{1},\mathbf{2}) = \chi_0(\mathbf{1},\mathbf{2}) + \int\!\text{d}(\mathbf{3}\mathbf{4}) \chi_0(\mathbf{1},\mathbf{3}) v(\mathbf{3},\mathbf{4}) \chi(\mathbf{4},\mathbf{2}) \, ,
\end{align}
with spectral representation~\cite{vanSetten2013, Golze2019}
\begin{align}
    \chi(\br,\br',\omega) = \sum_n 2  
    \frac{\Omexc}{\omega^2-\left(\Omexc\right)^2}
    \rho^{(n)}(\br) \rho^{(n)}(\br') \, , \label{eq-reducible_pol}
\end{align}
where $\Omexc$ are the excitation energies of the system and $\rho^{(n)}(\br)$ are transition densities of a given excitation index $n$.
The crucial aspect to accurately describe the electronic response is the pole structure of $\chi$ in Eq.~\eqref{eq-reducible_pol}, i.e.~inserting accurate excitation energies $\Omexc$.
These are typically computed from a Casida-like equation~\cite{vanSetten2013,Schambeck2024}
\begin{align}
    \left( \begin{array}{cc}\mathds{A} &  \mathds{B}\\\mathds{B} &  \mathds{A}\end{array} \right)\left( \begin{array}{cc}\bX^{(n)}\\\bY^{(n)}\end{array} \right) = \Omexc\left(\begin{array}{cc}\mathds{1}&0\\0&-\mathds{1}\end{array}\right)\left(\begin{array}{cc}\bX^{(n)}\\\bY^{(n)}\end{array}\right)  \, , \label{eq-ABBA}
\end{align}
where the entries of the eigenvectors $\Xian$ and $\Yian$ enter the transition densities $\rho^{(n)}(\br)$.
The specific entries of the matrices $\mathds{A}$ and $\mathds{B}$ read for singlet states excited from a closed-shell ground state~\cite{McLACHLAN1964,Oddershede1984,Bauernschmitt1996,Furche2000,Ullrich2011,Ljungberg2015,Jacquemin2016}
\begin{align}
    A_{ia,jb} &= (\varepsilon_a-\varepsilon_i)\delta_{ij}\delta_{ab} + 2
    v_{ia,jb} + F_{ij,ab} \,, \nonumber
    \\
    B_{ia,jb} &= 2 v_{ia,bj} + F_{ib,aj} \,. \label{eq-contributions_ABBA_app_a}
\end{align}
Here, the single-particle energies of the electronic system $\varepsilon_p$ enter as well the bare Coulomb interaction 
\begin{align}
    v_{pq,rs} &= \!\int \! \text{d}^3r \, \text{d}^3r' \psi_p(\br) \psi_q(\br) \frac{1}{|\br - \br'|} \psi_r(\br') \psi_s(\br') \label{eq-components_unscreened_coulomb_app_A}
\end{align}
which comes with a spin factor 2 (cf. App.~\ref{app-derivation_dyn_dip_pol_and_abba_eq}).
Both the energies $\varepsilon_p$ as well as the matrix elements $F_{pq,rs}$ depend on the chosen excited state method:
For the reducible polarizability $\chi$~\eqref{eq-reducible_pol} in $GW$, the excitation energies and transition densities are computed via the random-phase approximation (RPA), which uses $F_{pq,rs}=0$ in Eq.~\eqref{eq-contributions_ABBA_app_a}~\cite{Ullrich2011, vanSetten2013, Schambeck2024}.
In contrast, other excited state methods employ a non-zero $F_{pq,rs}$, e.g. TDDFT employs the exchange-correlation kernel $F_{pq,rs}\rightarrow f^\text{xc}_{pq,rs}$~\cite{Bauernschmitt1996, Furche2000, Ullrich2011}, which is in principle exact, but practical calculations employ approximations like  PBE~\cite{Perdew1996} or PBE0~\cite{Adamo1999}  there.
For the BSE, $F_{pq,rs}$ is the screened Coulomb interaction $-W_{pq,rs}$~\cite{Ljungberg2015, Jacquemin2016}, as we discuss in Sec.~\ref{sec-theory}.
In Table~\ref{tab-excitation_energies_TDDFT_RPA_BSE}, we further report the excitation energies of these methods (RPA, TDDFT and BSE) with varying KS starting point (PBE and PBE0) and refinement of the entering eigenenergies (KS, ev$GW_0$ and ev$GW$).
In particular, we observe that RPA tends to strongly overestimate the first excitation energy when starting from PBE0 ($\Omega^{(n=1)}_{\text{RPA@KS}}$ with PBE0) or ev$GW$ ($\Omega^{(n=1)}_{\text{RPA@ev}GW}$ for both PBE/PBE0), whereas $\Omega^{(n=1)}_\text{RPA@KS}$ based on PBE agrees better with the  results from the higher-level methods TDDFT or $GW$-BSE (cf. Ref.~\cite{Schambeck2024} for a detailed discussion).
Hence, in order to compute $\chi$~\eqref{eq-reducible_pol} with physically meaningful excitation energies, we conclude that $W_0$ instead of $W$ should result in improved accuracy when choosing one of the earlier discussed $GW$ schemes.
For further details on the $GW$ approximation and different (self-consistent) solution schemes, we refer to Refs.~\cite{Martin2016, Golze2019, Reining2018}.
}
\section{Derivation of the dynamical dipole polarizability and the defining matrix equation}\label{app-derivation_dyn_dip_pol_and_abba_eq}
In Eq.~\eqref{eq-BSE_polarizatbility}, we have introduced the tensor elements of the dynamical dipole polarizability~\cite{Ullrich2011, Bruneval2016, Jacquemin2016}
\begin{align}
    \alpha_{\mu\mu'}(\omega) 
    = - \sum_{n>0} \frac{2 \,\Omega^{(n)}\, d^{(n)}_{\mu} d^{(n)}_{\mu'}}{(\omega+i\eta)^2-\left(\Omega^{(n)}\right)^2} \, , \label{eq-BSE_polariz_appendix}
\end{align}
with~\cite{Ljungberg2015, Bruneval2016, Jacquemin2016, Liu2020}
\begin{align}
    d^{(n)}_{\mu} = \sqrt{2} \sum_{i,a} \mu_{ia} \left(X_{ia}^{(n)} + Y_{ia}^{(n)}\right), \nonumber
\end{align}
where $(\Omexc,\Xian,\Yian)$ are solutions of Eq.~\eqref{eq-BSE_ABBA}.
In the following, we derive the form of $\alpha_{\mu\mu'}(\omega)$ as well as the defining matrix equation in Eq.~\eqref{eq-BSE_ABBA}.
We start from the general definition of the dynamical dipole polarizability~\cite{Ullrich2011, Ljungberg2015, Onida2002}
\begin{align}
    \alpha_{\mu\mu'}(\omega) = - \int\!d^3rd^3r' r_\mu \chi(\br,\br',\omega)r_{\mu'}'
    \label{eq-dyn_dip_pol_from_red_pol}
\end{align}
where we have introduced the reducible polarizability $\chi(\br,\br',\omega)$ \reviewin{(cf. Eq.~\eqref{eq-reducible_pol})}.
In TDDFT, one can directly use the $\chi$ to arrive at Eq.~\eqref{eq-BSE_polariz_appendix}~\cite{Ullrich2011}.
However, within the BSE framework (and similar frameworks like, e.g., TDHF~\cite{McLACHLAN1964}), the retarded four-point polarizability $L(\mathbf{1},\mathbf{2},\mathbf{3},\mathbf{4})$ is needed, which relates to the two-point $\chi$ in time-domain via $\chi(\mathbf{1},\mathbf{2})=L(\mathbf{1},\mathbf{1^+},\mathbf{2},\mathbf{2})$.
We denote combined space-spin-time coordinates as $\mathbf{1}=\{\br_1,s_1,t_1\}$ \reviewin{(as in App.~\ref{app-gw_schemes})} and, in the following, combined space-spin coordinates as $1=\{\br_1,s_1\}$.
Further, we follow the convention of the main text regarding indexing, such that $p,q,r,s$ denote a generic KS orbital, whereas $i,j$ denote occupied, i.e.~$i,j\in[1,N_\text{occ}]$, and $a,b$ denote unoccupied orbitals, i.e.~$a,b\in[N_\text{occ}+1,N_\text{occ}+N_\text{empty}]$.

Changing to frequency domain, we find the Bethe-Salpeter equation for the four-point polarizability~\cite{Strinati1988, Rohlfing2000, Onida2002, Ljungberg2015, Bechstedt2015, Martin2016}
\begin{align}
    L(1,2,3,4;\omega) = 
    L^0&(1,2,3,4;\omega) \nonumber \\
    &+ \int\!d(5,6,7,8) \, L^0(1,2,5,6;\omega) \nonumber \\
    &\qquad K(5,6,7,8)\,L(7,8,3,4;\omega) \,. \label{eq-BSE_for_L}
\end{align}
with the non-interacting four-point polarizability $L^0(1,2,3,4;\omega)$ and the BSE kernel $K(5,6,7,8)$.
In the $GW$ approximation (cf. Eq.~\eqref{eq-GW}) and neglecting dynamical screening, i.e. $W(\omega=0)$, the kernel reads~\cite{Ljungberg2015, Blase2018, Blase2020}
\begin{align}
    K(1,2,3,4) = v(1,3)\delta(1,2)\delta(3,4)-W(1,2)\delta(1,3)\delta(2,4) \, .
    \label{eq-BSE_kernel}
\end{align}
For the following derivation, we bring the four-point polarizability into the basis of spin orbitals $\psi_p(1)=\psi_p(\br_1)\sigma_p(s_1)$ with quasiparticle orbitals $\psi_p(\br_1)$ (which we take to be the KS orbitals from Eq.~\eqref{eq-basis_set_expansion_KS_orb}) and spin function $\sigma_p(s_1)$, which can be either $\alpha(s_1)$ or $\beta(s_1)$:
\begin{align}
    L(1,2,3,4;\omega) = \sum_{pqrs} \psi_p(1)\psi_q(2)L_{pq,rs}(\omega)\psi_r(3)\psi_s(4) \, , \label{eq-4pt_polariz_from_realspace_to_matrix}
\end{align}
with matrix elements
\begin{align}
    L_{pq,rs}(\omega)  = \int\!d(1,2,3,4)\, \psi_p(1)\psi_q(2) L(1,2,3,4;\omega) \psi_r(3)\psi_s(4) \, .
\end{align}
Note that we skip all complex conjugates since the KS orbitals are real-valued.
By means of the residue theorem, the matrix elements of $L^0$ can be computed from the interacting Green's function $G$ (cf. Eq.~\eqref{eq-G}), which becomes diagonal in the basis of the KS orbitals~\cite{Onida2002, Ljungberg2015}:
\begin{align}
    L^0_{pq,rs} (\omega) = 
    \delta_{pr} \delta_{qs} \frac{f_p-f_q}{\omega - (\varepsilon_q^\evgwnod - \varepsilon_p^\evgwnod ) +i \eta } \, , \label{eq-4pt_polariz_L0}
\end{align}
where  $i\eta$ (with small $\eta$) is included, since we are focusing on the retarded polarizability needed for  optical responses~\mbox{\cite{Martin2016, Ljungberg2015}}.
We rewrite Eq.~\eqref{eq-BSE_for_L} in matrix form
\begin{align}
    \mathbf{L}(\omega) = (\mathbf{1}-\mathbf{L}^0(\omega)\mathbf{K})^{-1} \mathbf{L}^0(\omega) \, ,
\end{align}
where we denote all matrices with generic indices $p,q,r,s$ in bold.
After some matrix algebra, we obtain~\cite{Onida2002, Ljungberg2015}
\begin{align}
    \mathbf{L}(\omega) =
     \mathbf{M}^{-1}(\omega) \mathbf{F}
    \label{eq-full_4pt_pol_matrix_MF}
\end{align}
where  $\mathbf{M}$ and $\mathbf{F}$ have a block structure in the product space of occupied and unoccupied orbitals, i.e.~the rows and columns are ordered  via electron-hole ($ia,jb$), hole-electron ($ai,bj$), electron-electron ($ii',jj'$) and hole-hole ($aa',bb'$) sector:
\begin{align}
    \mathbf{M}(\omega) = 
    \begin{NiceArray}[first-row]{l \left\lgroup cccc\right\rgroup }
        & jb & bj & jj' & bb' \\
        ia & \tilde{\Omega}-\mathds{K} & -\mathds{K} & -\mathds{K} & -\mathds{K} \\
        ai &  \mathds{K} & \tilde{\Omega}+\mathds{K} & \mathds{K} & \mathds{K} \\
        ii' & 0&0 & \tilde{\Omega}  &0 \\
        aa' & 0& 0& 0& \tilde{\Omega} 
    \end{NiceArray}
    \, ,
\end{align}
and
\begin{align}
    \mathbf{F} = 
    \begin{NiceArray}[first-row]{l \left\lgroup cccc\right\rgroup }
        & jb & bj & jj' & bb' \\
        ia & \mathds{1} & 0 & 0 & 0 \\
        ai &  0 & -\mathds{1} & 0 & 0 \\
        ii' & 0&0 & 0  &0 \\
        aa' & 0& 0& 0& 0 
    \end{NiceArray}
    \, .
\end{align}
Correspondingly,  $\tilde{\Omega}$ are block matrices of the respective sector with the denominator of Eq.~\eqref{eq-4pt_polariz_L0} as matrix elements: 
\begin{align}
    \tilde{\Omega}_{pq,rs}(\omega) = (\omega+i\eta  +\varepsilon^\evgwnod_p - \varepsilon^\evgwnod_q )\delta_{pr} \delta_{qs} \,.
\end{align}
The triangular shape of $\mathbf{M}(\omega)$ and the block form of $\mathbf{F}$ in Eq.~\eqref{eq-full_4pt_pol_matrix_MF} reduce the number of non-zero entries of $\mathbf{L}(\omega)$ considerably, i.e.~to  the upper-left 2x2 block $\mathds{L}^{ia\,\oplus \,ai}(\omega)$~\cite{Onida2002}.
Until now, all quantities still comprise the spin degree of freedom.
However, even for a singlet closed-shell ground state, the single-particle excitation is not diagonal in the spins, which one immediately sees by explicitly writing down the spin indices of the upper left block of $\mathbf{M}$.
The spin structure of $-\tilde{\Omega}+\mathds{K}$, which follows from the contractions in Eqs.~\eqref{eq-BSE_kernel} and \eqref{eq-4pt_polariz_L0}, then reads
\begin{align}
    \begin{NiceArray}[first-row]{l \left\lgroup cccc\right\rgroup }
        & \bar{\alpha}\bar{\alpha}' & \bar{\beta}\bar{\beta}' & \bar{\alpha}\bar{\beta} & \bar{\beta}\bar{\alpha} \\
        \alpha\alpha' & -\tilde{\Omega} + \mathds{V} -\mathds{W} & \mathds{V} & 0 & 0 \\
        \beta\beta' &  \mathds{V} & -\tilde{\Omega} + \mathds{V} -\mathds{W} & 0 & 0 \\
        \alpha\beta & 0&0 & -\tilde{\Omega} -\mathds{W}  &0 \\
        \beta\alpha & 0& 0& 0& -\tilde{\Omega}  -\mathds{W} 
    \end{NiceArray}
\end{align}
with  elements of the block matrices $\mathds{V}$ and $\mathds{W}$ given by $v_{pq,rs}$ and $W_{pq,rs}$ (cf.~Eqs.~\eqref{eq-components_unscreened_coulomb} and \eqref{eq-components_screened_coulomb}).
Bringing the upper block to diagonal form by the unitary transformation 
\begin{align}
    \frac{1}{\sqrt{2}}
    \begin{pmatrix}
        \mathds{1} &  \mathds{1} \\
        \mathds{1} & -\mathds{1}
    \end{pmatrix} \label{eq-spin_trafo}
    \, ,
\end{align}
i.e.~transforming the matrix to the basis of singlet and triplet solutions, we obtain the singlet-triplet factor (with $\alpha^\mathrm{(S)}=2$ and $\alpha^\mathrm{(T)}=0$) in Eq.~\eqref{eq-BSE_ingredients}~\cite{Ljungberg2015, Bechstedt2015, McLACHLAN1964}\footnote{Since Ref.~\cite{McLACHLAN1964} has derived these matrix elements for TDHF, the $W_{pq,rs}$ are replaced by $v_{pq,rs}$ there.}
\begin{align}
    A_{ia,jb} &= (\varepsilon_a^{\evgwnod}-\varepsilon_i^{\evgwnod})\delta_{ij}\delta_{ab} + \alpha^\mathrm{(S/T)}
    v_{ia,jb} - W_{ij,ab} \,, \nonumber
    \\
    B_{ia,jb} &= \alpha^\mathrm{(S/T)} v_{ia,jb} - W_{ib,aj} \,.
    \nonumber
\end{align}
With this short-hand notation and by further utilizing the symmetry of the kernel $K_{pq,rs}$ under index-permutation, we can rewrite the upper left 2x2 block of $\mathbf{M}$ as
\begin{align}
    \mathds{M}^{ia\,\oplus \,ai}(\omega) = -
    \begin{pmatrix}
        \mathds{1} & 0 \\
        0 & - \mathds{1}
    \end{pmatrix}
    \begin{pmatrix}
        \mathds{A}-(\omega+i\eta)\mathds{1} & \mathds{B} \\
        \mathds{B} & \mathds{A}+(\omega+i\eta)\mathds{1}
    \end{pmatrix}
    \, ,
\end{align}
and thus for the corresponding block of $\mathbf{L}$~\cite{Jacquemin2016}
\begin{align}
    \mathds{L}^{ia\,\oplus \,ai}(\omega) = - 
    \left[
    \begin{pmatrix}
        \mathds{A} & \mathds{B} \\
        \mathds{B} & \mathds{A}
    \end{pmatrix}
    - (\omega + i\eta)
    \begin{pmatrix}
        \mathds{1} & 0 \\
        0 & \mathds{-1}
    \end{pmatrix}
    \right]^{-1}
    \, . \label{eq-4pt_polarizab_part_hole}
\end{align}
We do not explicitly denote the spin here, but the $\alpha^\text{(S,T)}$ in $\mathds{A}$ and $\mathds{B}$ implicitly accounts for the spin degrees of freedom.

\allowdisplaybreaks[1] 
In the next step, we  use the spectral representation of Eq.~\eqref{eq-4pt_polarizab_part_hole} (cf. Appendix of Ref.~\cite{Furche2001}) to rewrite $\mathds{L}^{ia\,\oplus \,ai}(\omega)$.
Thus, Eq.~\eqref{eq-4pt_polarizab_part_hole} also defines the central eigenvalue equation in Eq.~\eqref{eq-BSE_ABBA}.
With the spectral representation, we obtain four blocks with elements
\begin{subequations}  \vspace{-1em}  \begin{align}
        L_{ia,jb}(\omega) &= \sum_{n=1}^{N_\text{occ}N_\text{empty}} \frac{\Xian \Xjbn}{\omega+i\eta - \Omexc} - \frac{\Yian \Yjbn}{\omega+i\eta + \Omexc} \,, \\
        L_{ia,bj}(\omega) &= \sum_{n=1}^{N_\text{occ}N_\text{empty}} \frac{\Xian \Yjbn}{\omega+i\eta - \Omexc} - \frac{\Yian \Xjbn}{\omega+i\eta + \Omexc} \,, \\
        L_{ai,jb}(\omega) &= \sum_{n=1}^{N_\text{occ}N_\text{empty}} \frac{\Yian \Xjbn}{\omega+i\eta - \Omexc} - \frac{\Xian \Yjbn}{\omega+i\eta + \Omexc} \,, \\
        L_{ai,bj}(\omega) &= \sum_{n=1}^{N_\text{occ}N_\text{empty}} \frac{\Yian \Yjbn}{\omega+i\eta - \Omexc} - \frac{\Xian \Xjbn}{\omega+i\eta + \Omexc} \, .    \end{align}\label{eq_block_struct_of_4pt_polarizab}
\end{subequations}
In the case of the TDA, all components comprising an element of $\mathbf{Y}^{(n)}$ drop out and we recover Eq.~(32) from Ref.~\cite{Ljungberg2015}.
Coming back to the dynamical dipole polarizability $\alpha_{\mu\mu'}(\omega)$, we write Eq.~\eqref{eq-dyn_dip_pol_from_red_pol} including the spin channels as
\begin{align}
    \alpha_{\mu\mu'}(\omega) = - \sum_{s_1 s_2} \int\!d^3r_1\,d^3r_2 \, r^{(\mu)}_1 L(1,1,2,2,\omega)r_2^{(\mu)} \, .
\end{align}
With Eq.~\eqref{eq-4pt_polariz_from_realspace_to_matrix} and after transforming to the singlet-triplet basis (cf. Eq.~\eqref{eq-spin_trafo}), only the singlet contribution remains with a factor of $\sqrt{2}$ resulting from the sum over spin indices~\cite{Bechstedt2015, Ljungberg2015, Jacquemin2016, McLACHLAN1964}:
\begin{align}
    \alpha_{\mu\mu'}(\omega) = - (\sqrt{2})^2 \sum_{pqrs} \mu_{pq} \mu'_{rs} L_{pqrs}^\text{(Singlet)}(\omega) \, . \label{eq-dyn_dip_pol_from_4pt}
\end{align}
Combining Eq.~\eqref{eq-dyn_dip_pol_from_4pt} with the block structure of $L_{pq,rs}(\omega)$ in Eqs.~\eqref{eq_block_struct_of_4pt_polarizab}, we arrive at Eq.~\eqref{eq-BSE_polariz_appendix}, where the spin factor is absorbed in the transition dipole moments $d_\mu^{(n)}$ from Eq.~\eqref{eq-BSE_trans_mom}.
%

\section{Energy cutoff for the BSE}\label{app-cutoff_checks}
\begin{figure}
    \centering
    \includegraphics[width=\linewidth]{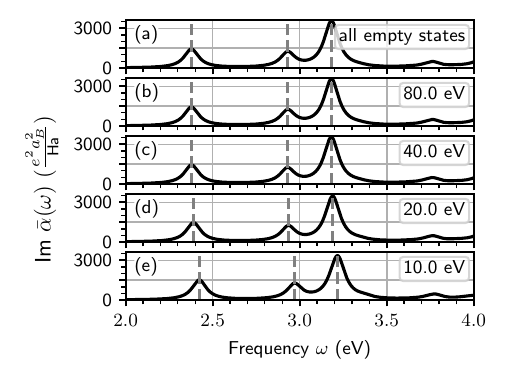}
    \caption{
    Optical absorption spectra computed from $\mathrm{Im}\ \bar{\alpha}(\omega)$ [Eq.~\eqref{eq-spatial_avg_polarizability}] for all unoccupied states (a) and four different cutoffs $E_\mathrm{cut}^\mathrm{empty}$(b-e) with a broadening of $\eta=0.05$~eV \reviewin{for a nanographene with $L=4$}.
    The changes for the first three peaks (grey vertical lines) are reported in Table~\ref{tab-cutoff_checks_spectrum}.
    }
    \label{fig-cutoff_checks_nanographene}
\end{figure}

In the following, we introduce the details about the truncation of the number of orbitals, which enter the BSE in Eq.~\eqref{eq-BSE_ABBA}.
In the following, we discuss the truncation of the considered number of orbitals in Eq.~\eqref{eq-BSE_ABBA}, motivated by its heavy computational scaling:
Since both the number of occupied $N_\mathrm{occ}$ and unoccupied orbitals $N_\mathrm{empty}$ scales linearly with the system size $N$, the index structure of Eq.~\eqref{eq-BSE_ingredients} leads to a memory scaling of $\mathcal{O}(N_\mathrm{occ}^2N_\mathrm{empty}^2) = \mathcal{O}(N^4)$ and a runtime scaling of $\mathcal{O}(N_\mathrm{occ}^3N_\mathrm{empty}^3)=\mathcal{O}(N^6)$.
Beyond that, the computational effort associated with the calculation of the $GW$ quasiparticle energies needed for Eq.~\eqref{eq-BSE_ingredients} scales with the number of included orbital.
Hence, we adopt the idea of Ref.~\cite{Liu2020} to impose an energy cutoff on the included states since we do not expect that low-lying optical excitations are strongly affected by the high-lying states, which reduces the number of states and therefore enables calculations with a reasonable amount of computational resources.
Specifically, we apply the truncation as follows:
(i) We already truncate the number of orbitals based on their KS energies, which are refined in the $GW$-module (cf. Fig.~\ref{fig-flowchart}) to keep the computational effort to the possible minimum. 
Correspondingly, only these refined states enter the construction of Eq.~\eqref{eq-BSE_ingredients} (cf. Fig.~\ref{fig-flowchart}).
(ii) In addition to the cutoff for unoccupied states, we also specify a separate cutoff for the occupied states targeting core levels.
We test the accuracy of this cutoff procedure for a nanographene with $L=4$.
In order to only exclude the core level states, we choose a cutoff for the occupied states of $E_\mathrm{cut}^\mathrm{occ}=80$~eV, which keeps excitation energies virtually unchanged (within less than 1~meV), but reduces $N_\mathrm{occ}$ by approximately $30$\% from 180 to 124.
Given the negligible influence of $E_\mathrm{cut}^\mathrm{occ}=80$~eV on the accuracy of the calculation, we use it throughout all further cutoff checks.
Since the unoccupied states exhibit a continous distribution of orbital energies $\epsilon_a$, we check four different cutoff values $E_\mathrm{cut}^\mathrm{empty}=10,20,40,80$~eV and plot the resulting spectra in Fig.~\ref{fig-cutoff_checks_nanographene}.
We observe only minor changes in the absorption spectrum, even for the smallest cutoff of $E_\mathrm{cut}^\mathrm{empty}=10$~eV, which reduces unoccupied states from 1195 to 204.
\reviewin{In Table~\ref{tab-Number_empty_orbitals}, we additionally report the respective number of unoccupied states for each of the cutoffs $E_\mathrm{cut}^\mathrm{empty}$ in Fig.~\ref{fig-cutoff_checks_nanographene}.}
\begin{table}
    \centering
    \begin{tabular}{w{l}{1.5cm} | w{r}{1cm} | w{r}{1cm} | w{r}{1cm} | w{r}{1cm} | w{r}{1cm}}
         $E_\mathrm{cut}^\mathrm{empty}$
         & all
         & 80~eV
         & 40~eV
         & 20~eV
         & 10~eV 
         \\[0.1em] \hline\\[-0.8em] 
         $N_\text{empty}$
         & 1195
         & 1138
         & 709
         & 425
         & 204
    \end{tabular}
    \caption{
    \reviewin{Number of unoccupied states $N_\text{empty}$ for a nanographene with $L=4$ resulting from the cutoffs $E_\mathrm{cut}^\mathrm{empty}$ in Fig.~\ref{fig-cutoff_checks_nanographene}.}
    }
    \label{tab-Number_empty_orbitals}
\end{table}

\begin{table}
    \centering
    \begin{tabular}{l|c c|c|rr}
         $E_\mathrm{cut}^\mathrm{empty}$
         & $\Omega^{(p)}$
         & $\bar{\alpha}^{(p)}$
         & $\Delta \Omega^{(p)}$
         & ${\Delta  \Omega^{(p)}}/{ \Omega^{(p)}}$
         & ${\Delta \bar{\alpha}^{(p)}}/{\bar{\alpha}^{(p)}}$ \\[-0.3em] 
         \quad (eV)
         &\quad (eV)
         &\quad $(\frac{ e^2 a_B^2}{ \mathsf{Ha} })$
         &\quad (meV)
         &\quad (\%)
         &\quad (\%) \\
\midrule
 & 2.378 & 1.954 & - & - & - \\
Ref. & 2.933 & 1.709 & - & - & - \\
 & 3.183 & 4.828 & - & - & - 
\\[0.3em] \hline\\[-0.8em] 
 & 2.378 & 1.954 & 0 & 0.0 & 0.0 \\
80.0 & 2.933 & 1.709 & 0 & 0.0 & 0.0 \\
 & 3.183 & 4.827 & 0 & 0.0 & 0.0 
\\[0.3em] \hline\\[-0.8em] 
 & 2.379 & 1.963 & 1 & 0.0 & 0.4 \\
40.0 & 2.931 & 1.707 & -2 & -0.1 & -0.1 \\
 & 3.182 & 4.803 & -1 & 0.0 & -0.5 
\\[0.3em] \hline\\[-0.8em] 
 & 2.390 & 1.995 & 12 & 0.5 & 2.1 \\
20.0 & 2.936 & 1.701 & 4 & 0.1 & -0.4 \\
 & 3.190 & 4.699 & 7 & 0.2 & -2.7 
\\[0.3em] \hline\\[-0.8em] 
 & 2.423 & 2.064 & 45 & 1.9 & 5.6 \\
10.0 & 2.972 & 1.742 & 39 & 1.3 & 2.0 \\
 & 3.217 & 4.534 & 35 & 1.1 & -6.1 \\
    \end{tabular}
    \caption{
    Peak frequencies $\Omega^{(p)}$ and heights $\bar{\alpha}^{(p)}$ as well as the respective deviations $\Delta \Omega^{(p)}$, $\Delta \Omega^{(p)}/\Omega^{(p)}$, and ${\Delta \bar{\alpha}^{(p)}}/{\bar{\alpha}^{(p)}}$ of the three highest peaks $p=1,2,3$ extracted from Fig.~\ref{fig-cutoff_checks_nanographene} for four  cutoffs $E_\mathrm{cut}^\mathrm{empty}$.
    We do not explicitly denote the imaginary parts for the the reported peak heights for the sake of readability.
    }
    \label{tab-cutoff_checks_spectrum}
\end{table}
In Table~\ref{tab-cutoff_checks_spectrum}, we report the extracted peak positions $\Omega^{(p)}$ and heights $\bar{\alpha}^{(p)}$ from the first three ($p=1,2,3$) peaks in Fig.~\ref{fig-cutoff_checks_nanographene}.
Even for the smallest cutoff $E_\mathrm{cut}^\mathrm{empty}=10$~eV, we only observe minor changes of less than $50$~meV in $\Omega^{(p)}$ and 10\% in the relative peak heights ${\Delta \bar{\alpha}^{(p)}}/{\bar{\alpha}^{(p)}}$, respectively.
Beyond the absorption spectrum, we report the impact of the cutoffs on the spatial descriptors, specifically on the longitudinal size of the excitation $d_\mathrm{exc}^{(x)}$ in Table~\ref{tab-cutoff_checks_descriptors}.
We determine the corresponding excitation level $n$ from the largest oscillator strength $f^{(n)}$ (corresponding to the highest peak of $\mathrm{Im}\ \bar{\alpha}(\omega)$, cf. Eq.~\eqref{eq-absorption_peaks_from_theory_as_epsilon}) for each cutoff value $E_\mathrm{cut}^\mathrm{empty}$.
Again, we find only minor deviations around 2\%.
\reviewin{
Based on these results for $L=4$, we execute the calculations for all $L$ in the main text employing the cutoffs $E_\mathrm{cut}^\mathrm{occ}=80$~eV and $E_\mathrm{cut}^\mathrm{empty}=10$~eV.
Our motivation to rely on the explicit energy cutoffs rather than the number of states (cf. Table~\ref{tab-Number_empty_orbitals}) is based on a rough estimate via first-order perturbation theory, which estimates the contribution of higher-lying unoccupied states based on their respective level energy.
This enables simulation of $GW$-BSE for large system sizes with up to 242 atoms for $L=13$, which in turn corresponds to $N_\text{occ}=394$ and $N_\text{empty}=573$, i.e.~the BSE matrices $\mathds{A},\mathds{B}$ in Eq.~\eqref{eq-BSE_ABBA} have dimension $N_\text{occ}N_\text{empty}=2.3\cdot 10^5$.
}
\reviewout{
Taking into account the sizable acceleration provided by these cutoffs, we execute the calculations in the main text employing the cutoffs $E_\mathrm{cut}^\mathrm{occ}=80$~eV and $E_\mathrm{cut}^\mathrm{empty}=10$~eV, enabling simulation of $GW$-BSE for large system sizes with up to 242 atoms for $L=13$, which then corresponds to $N_\text{occ}N_\text{empty}=2.3\cdot 10^5$.
}

\begin{table}
    \centering
\begin{tabular}{l|cc|rr}
 $E_\mathrm{cut}^\mathrm{empty}$ & $\Omega^{(n)}$~(eV) & $d_\mathrm{exc}^{(x)}$~(\AA) & $\Delta \Omega^{(n)}$~(eV) & $\frac{\Delta d_\mathrm{exc}^{(x)}}{d_\mathrm{exc, ref}^{(x)}}$~(\%) \\
\midrule
Ref. & 3.184 & 5.813 & - & - \\
80.0 & 3.184 & 5.811 & 0.000 & 0.0 \\
40.0 & 3.184 & 5.758 & 0.000 & -0.9 \\
20.0 & 3.193 & 5.704 & 0.009 & -1.9 \\
10.0 & 3.223 & 5.679 & 0.038 & -2.3 \\
\end{tabular}
    \caption{Peak frequencies $\Omega^{(n)}$ and corresponding size of the excitation in the longitudinal direction $d_\mathrm{exc}^{(x)}$ as well as the respective deviations $\Delta \Omega^{(p)}$ and ${\Delta d_\mathrm{exc}^{(x)}}/{d_\mathrm{exc, ref}^{(x)}}$ for the excitation level $n$ with the largest oscillator strength $f^{(n)}$ for four explicit cutoffs $E_\mathrm{cut}^\mathrm{empty}$.
    }
    \label{tab-cutoff_checks_descriptors}
\end{table}

\section{Isotropic polarizability and photoabsorption cross section}\label{app-key_quantitities_for_opt_exc}
In the main text and in App.~\ref{app-derivation_dyn_dip_pol_and_abba_eq}, we have derived a form for the dynamical dipole polarizability tensor $\alpha_{\mu \mu'}(\omega)$, which enables the computation of  optical absorption spectra with $GW$-BSE.
In the following, we introduce a number of additional quantities, which can be used to describe optical excitations and are computed by our implementation.
In gases and liquids, where molecules are oriented randomly, the complicated tensor structure of $\alpha_{\mu \mu'}(\omega)$ can be simplified by averaging over the three cartesian components, leading to the spatial average
\begin{align}
    \bar{\alpha}(\omega) = \frac{1}{3} \sum_{\mu\in\{x,y,z\}} \alpha_{\mu,\mu}(\omega) \,. \label{eq-spatial_avg_polarizability}
\end{align}
\reviewout{Naturally, these defines the oscillator strengths $f^{(n)}$}
We can then rewrite $\bar{\alpha}(\omega)$,
\begin{align}
    \bar{\alpha}(\omega)
    =
    -  \mathrm{Im}\left[
      \sum_{n>0} \frac{f^{(n)}}{(\omega+i\eta)^2-\left(\Omega^{(n)}\right)^2}
      \right]
    \, ,\label{eq-BSE_avg_spectrum}
\end{align}
where the residues are the oscillator strengths $f^{(n)}$~\cite{Jacquemin2016,Liu2020}
\begin{align}
    f^{(n)} = \frac{2}{3}\, \Omega^{(n)} \sum_{\mu\in\{x,y,z\}} | d^{(n)}_{\mu} |^2
    \, , \label{eq-BSE_oscillator_strength}
\end{align}
Here, we have used the transition dipole moments from Eq.~\eqref{eq-BSE_trans_mom}
\begin{align}
    d^{(n)}_{\mu} = \sqrt{2} \sum_{i,a} \langle \psi_i|\hat{\mu}| \psi_a \rangle \left(X_{ia}^{(n)} + Y_{ia}^{(n)}\right).
    \nonumber
\end{align}
We further compute the photoabsorption cross section tensor $\sigma_{\mu\mu'}(\omega)$, which is defined by~\cite{Ullrich2011} 
\begin{align}
    \sigma_{\mu\mu'}(\omega)  = \frac{4\pi \omega}{c} \text{Im} \left[ \alpha_{\mu\mu'}(\omega) \right] \, . \label{eq-crosssection_and_polarizability}
\end{align}
\section{Computational details of \textit{GW}-BSE and TDDFT calculations}\label{app-computational_details}
The majority of the calculations in this paper is done with the CP2K package~\cite{CP2K,Kuhne2020,Iannuzzi2026}.
CP2K uses Gaussian basis sets for representing the KS orbitals in Eq.~\eqref{eq-basis_set_expansion_KS_orb}, where we employ the aug-cc-pVDZ basis set~\cite{Dunning1989,Kendall1992} throughout the paper, provided by the EMSL database~\cite{Pritchard2019}.
These all-electron calculations are executed within the Gaussian and augmented plane-waves scheme~\cite{Lippert1999}.
The $GW$ implementation is based on Ref.~\cite{Wilhelm2016}, where the the analytic continuation is applied in the computation of the self-energy~\eqref{eq-GW} using 16 Padé approximants and a frequency grid with 100 points unless stated otherwise.
By using ev$GW_0$@PBE, we get rid of numerical artifacts caused by the pole structure of the self-energy~\cite{Veril2018, Schambeck2024}, which can occur for the PBE functional~\cite{Perdew1996}.
The reference calculations to check the numerical validity of our implementation have been done using the FHI-aims code~\cite{Blum2009, Ren2012, Liu2020, Abbott2025}, where we have used the \textit{gaussian-tight} settings together with the aug-cc-pVDZ basis set~\cite{Dunning1989,Kendall1992} applying the same parameters for the ev$GW_0$ calculations specified before.
The TDDFT module in CP2K is based on the Tamm Dancoff approximation~\cite{Strand2019, Hehn2022, SertcanGokmen2024, Vogt2025, Vogt2025a}, which uses the auxiliary density matrix method to enable efficient calculations with hybrid functionals.
There, we have chosen the aug-cc-pVDZ~\cite{Dunning1989,Kendall1992} and aug-DZVP-MOLOPT-GTH-tier-2 basis sets (basis set generation recipe as in Refs.~\cite{Pasquier2025, Pasquier2026}) in combination with the admm-2~\cite{Kumar2018} auxiliary basis.
The exchange-correlation functionals  employed in this work (cf. Sec.~\ref{sec-spatial_descriptors} and App.~\ref{app-tddft_with_pbeh}) are~PBE~\cite{Perdew1996}, BLYP~\cite{Becke1988, Lee1988}, PBE0~\cite{Adamo1999}, B3LYP~\cite{Lee1988, Becke1993, Stephens1994}, HSE06~\cite{Heyd2003, Krukau2006} as well as CAM-B3LYP~\cite{Yanai2004}.
%

\section{\textit{GW}-BSE benchmark on Thiel's set}\label{app-thiels_set}
\begin{figure}
    \centering
    \includegraphics[width=3.375in]{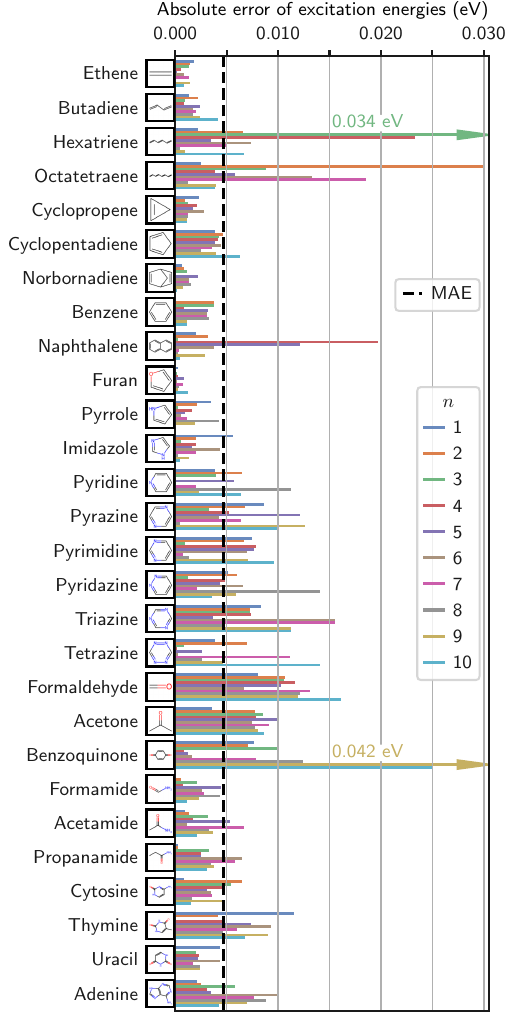}
    \caption{
    Absolute error~$|\Omega^{(n)}_\mathrm{CP2K}-\Omega^{(n)}_\mathrm{aims}|$ of BSE excitation energies $\Omexc$ computed from CP2K and FHI-aims by solving Eq.~\eqref{eq-BSE_ABBA} with BSE@ev$GW_0$@PBE. 
    The analytical continuation for the $GW$-modules in both codes was applied with 16 parameters of the Padé function and 100 points for the frequency grid, which results in an mean absolute error (MAE) of 4.7~meV and a maximum absolute error of 42.3~meV over the ten lowest excitation energies across all molecules.
    }
    \label{fig-Thiels_set_with_AC_16_100}
\end{figure}

In this appendix, we provide further details about the parameters of the analytic continuation, which we have used to compute the real-frequency self-energy both in the $GW$-modules of CP2K~\cite{Wilhelm2016} and FHI-aims~\cite{Blum2009,Ren2012}.
In Fig.~\ref{fig-thiels_set_128+500}, we have presented a benchmark for \textit{Thiel's set}~\cite{Schreiber2008}, where the CP2K calculation employed 128 parameters for the Padé function and 500 points in the frequency grid whereas we have used 16 and 100, respectively, for the calculation with FHI-aims, resulting in mean absolute error (MAE; cf. Eq.~\eqref{eq-MAE_CP2K_aims}) of 2.7~meV and a maximum absolute error of 30.2~meV.
In Fig.~\ref{fig-Thiels_set_with_AC_16_100}, we show another comparison for 16 Padé parameters and 100 points for the frequency grid in both CP2K and FHI-aims.
We observe slightly larger deviations between the implementations of 4.7~meV MAE and a maximum absolute error of 42.3~meV over the ten lowest excitation energies across all molecules.

\section{Absorption spectra from experiment and theory:  dielectric function and polarizability}
\label{app-diel_func_from_pol}
In this appendix, we connect the optical absorption spectrum of a rectangular nanographene from experiment~\cite{Denk2014} quantified via the dielectric function and from $GW$-BSE computed via the polarizability of a finite nanographene.
To that end, we start from the experimental reflectance difference measurements, from which they extract the anisotropic dielectric function $\Delta\epsilon = \epsilon_x - \epsilon_y$.
We approximate the anisotropic dielectric function $\Delta\epsilon$ by its longitudinal component $\epsilon_x$, i.e.~$\Delta\epsilon \approx \epsilon_x$, since the overall optical spectrum is dominated by the longitudinal component (cf. Ref.~\cite{Denk2014} and also Sec.~\ref{sec-spectra_of_nanographenes_theory}).
The dielectric function along the long ribbon axis $\epsilon_x^\text{(exp.)}$ is then obtained from the three Lorentzian peaks ($n=1,2,3$) at frequencies $\Omega^{(n)}_\text{(exp.)}$ reported in the supporting information of Ref.~\cite{Denk2014}:
\begin{align}
    \epsilon_x^\text{(exp.)} (\omega) = \sum_{n=1}^3 \frac{A_\text{(exp.)}^{(n)}}{{(\Omega_\text{(exp.)}^{(n)}})^2 - \omega^2 - i \omega \Gamma_\text{(exp.)}^{(n)}} \, , \label{eq-lorentzians_from_experiment}
\end{align}
where $A^{(n)}_\text{(exp.)}$ and $\Gamma^{(n)}_\text{(exp.)}$ further denote the amplitude and the width of the respective transition $p$, respectively.
In our analysis, the amplitudes $A^{(n)}_\text{(exp.)}$ are expressed with the dimension $(\text{eV})^2$, ensuring that the resulting dielectric function $\epsilon_x^\text{(exp.)}$ reproduces the dimensionless lineshape $\Delta\epsilon^\text{(exp.)} \approx \epsilon^\text{(exp.)}_x$ from Ref.~\cite{Denk2014}.
In Fig.~\ref{fig-abs_spectra_BSE_vs_exp}(a), we plot the imaginary part $\text{Im}\ \epsilon_x(\omega)$ as well as the height of an individual transition $p$:
\begin{align}
    \mathrm{Im}\ \epsilon_x^\text{(exp.)} (\omega=\Omega^{(n)}) = \frac{A_\text{(exp.)}^{(n)} }{\Omega_\text{(exp.)}^{(n)} \Gamma_\text{(exp.)}^{(n)}} \, . \label{eq-absorption_peaks_from_exp}
\end{align}
On the side of $GW$-BSE, we utilize Eq.~\eqref{eq-relation_prop_epsilon_alpha} and plot $\text{Im}\ \alpha^{(L)}_{xx}(\omega)$ with a broadening of $\eta = 0.3$~eV in Fig.~\ref{fig-abs_spectra_BSE_vs_exp}(b), for which we obtain a similar lineshape as in the experimental data.
Additionally, we show the height of an individual transition $n$ obtained from the BSE (cf. Eq.~\eqref{eq-BSE_polarizatbility}) 
\begin{align}
    \mathrm{Im}\ \alpha_{xx} (\omega=\Omega^{(n)}) =  f^{(n)}_x\frac{2\eta \Omega^{(n)}}{4\eta^2(\Omega^{(n)})^2 + \eta^4}
    \, , \label{eq-absorption_peaks_from_theory_as_epsilon}
\end{align}
where we abbreviate the \textit{oscillator strength in $x$-direction} as $f^{(n)}_x = 2 \Omega^{(n)} |d_x^{(n)}|^2$ (cf. Eq.~\eqref{eq-BSE_oscillator_strength}).

\section{Effect of the Tamm-Dancoff approximation in \textit{GW}-BSE}\label{app-bse_abba_vs_TDDFT_tda}
\begin{figure}
    \centering
    \includegraphics[width=\linewidth]{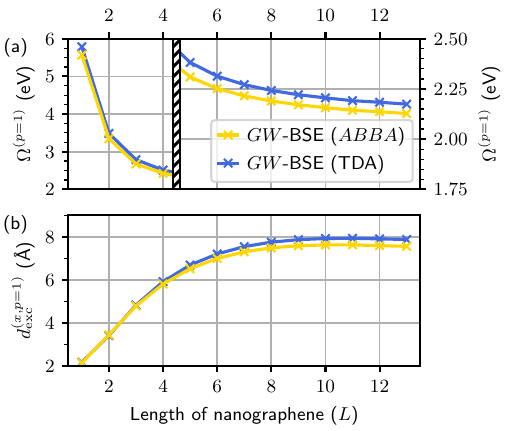}
    \caption{
    Excitation energy $\Omega^\text{(p=1)}$ and size of the excitation $d_\mathrm{exc}^{(x,\text{p=1})}$ computed from BSE@ev$GW_0$ via Eq.~\eqref{eq-BSE_ABBA} (yellow) and Eq.~\eqref{eq-TDA_of_BSE} (blue).
    (a) Peak frequency of the first prominent peak in the optical absorption spectrum $\Omega^\text{(p=1)}$.
    (b) Longitudinal size of the excitation for the first bright peak $d_\mathrm{exc}^{(x,\text{p=1})}$, Eq.~\eqref{eq-longitudinal_excitation_size}.
    }
    \label{fig-bse_abba_vs_tda_peaks_and_descr}
\end{figure}

\begin{figure}
    \centering
    \includegraphics[width=\linewidth]{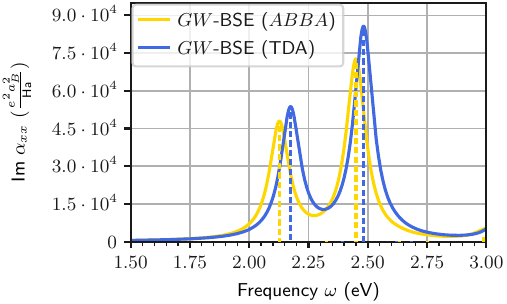}
    \caption{
    Optical absorption spectrum  for $GW$-BSE via Eq.~\eqref{eq-BSE_ABBA} (yellow) and Eq.~\eqref{eq-TDA_of_BSE} (blue) from $\mathrm{Im}\ \alpha_{xx}(\omega)$ [Eq.~\eqref{eq-BSE_polarizatbility}].
    Dashed lines  denote the contribution of individual excitations $n$ to $ \mathrm{Im}\ \alpha_{xx}(\omega=\Omega^{(n)})$.
    }
    \label{fig-bse_abba_vs_tda_spectra}
\end{figure}

In Fig.~\ref{fig-tddft_gga_hyb_rs_hyb_vs_BSE}, we have compared the excitation properties computed from BSE@ev$GW_0$ against TDDFT with different exchange–correlation functionals, where we have employed the TDA [Eq.~\eqref{eq-TDA_of_BSE}] for both frameworks.
Here, we want to assess the deviations introduced by the TDA and compare the results obtained from the TDA in Eq.~\eqref{eq-TDA_of_BSE} against the solution of the full $ABBA$-matrix in Eq.~\eqref{eq-BSE_ABBA} within BSE@ev$GW_0$.
In Fig.~\ref{fig-bse_abba_vs_tda_peaks_and_descr}, we plot the excitation energy and the corresponding size of the respective lowest bright state ($p=1$) for the solutions of Eqs.~\eqref{eq-BSE_ABBA} and \eqref{eq-TDA_of_BSE}.
We observe only minor deviations of at most $0.22$~eV for energies (at small $L$) and of $0.32$~\AA\ in sizes of the excitation (at large $L$) for all nanographene lengths $L$.
The details of the respective optical absorption spectra computed via Eq.~\eqref{eq-BSE_polarizatbility} are shown in Fig.~\ref{fig-bse_abba_vs_tda_spectra} for $L=13$.
Despite minor changes in peak frequencies and heights, we conclude that our findings from Sec.~\ref{sec-spatial_descriptors} do not change when diagonalizing Eq.~\eqref{eq-BSE_ABBA} instead of Eq.~\eqref{eq-TDA_of_BSE}.

\section{TDDFT with manually tuned PBEh}\label{app-tddft_with_pbeh}
\begin{figure}
    \centering
    \includegraphics[width=\linewidth]{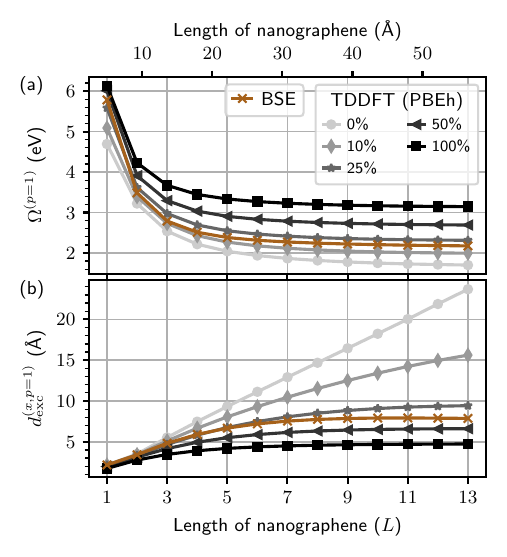}
    \caption{
    Excitation energy $\Omega^\text{(p=1)}$ and size of the excitation $d_\mathrm{exc}^{(x,\text{p=1})}$ computed from BSE@ev$GW_0$ and TDDFT (both in TDA; effect of TDA investigated in App.~\ref{app-bse_abba_vs_TDDFT_tda}) employing the PBEh functional with varying amount of exact exchange (exact exchange fraction shown in the legend).
    (a) Peak frequency of the first prominent peak in the optical absorption spectrum $\Omega^\text{(p=1)}$.
    (b) Longitudinal size of the excitation for the first bright peak $d_\mathrm{exc}^{(x,\text{p=1})}$, Eq.~\eqref{eq-longitudinal_excitation_size}.
    }
    \label{fig-tddft_appendix_pbeh_descriptors_and_peaks}
\end{figure}
In Sec.~\ref{sec-spatial_descriptors}, we have discussed several popular exchange–correlation functionals and the respective properties of the first bright excitation $p=1$.
There, we have observed that the fraction of exact exchange in the exchange–correlation functional in TDDFT strongly influences both the excitation energy and the size of the excitation.  
However, the employed functionals do not only feature different fraction of exact exchange, but also different contributions for the underlying exchange and correlation functionals.
In order to unambiguously tie the accuracy for predictions of excitation properties to the exact exchange fraction, we show energies and sizes of the first bright excitation for varying amount of exact exchange in the PBEh functional in Fig.~\ref{fig-tddft_appendix_pbeh_descriptors_and_peaks}, i.e.~PBE0 (25\%)  with exact exchange varied between 0\% (PBE) and 100\%.
Again, we observe that both excitation energies and sizes critically depend on the chosen exchange exact fraction.

\newpage

\bibliography{Bibliothek}
\end{document}